\documentclass[aps,groupedaddress,floatfix,preprintnumbers,nofootinbib,eqsecnum]{revtex4}
\usepackage{amssymb,amsmath,graphicx,multirow}

\begin{document}

\preprint{UH511-1191-12}

\title{Distinguishing Dynamical Dark Matter at the LHC}
\author{Keith R. Dienes$^{1,2,3}$\footnote{E-mail address:  {\tt dienes@physics.arizona.edu}},
      Shufang Su$^{3}$\footnote{E-mail address:  {\tt shufang@physics.arizona.edu}},  
      Brooks Thomas$^{4}$\footnote{E-mail address:  {\tt thomasbd@phys.hawaii.edu}}}
\affiliation{
     $^1$ Physics Division, National Science Foundation, Arlington, VA  22230  USA\\
     $^2$ Department of Physics, University of Maryland, College Park, MD  20742  USA\\
     $^3$ Department of Physics, University of Arizona, Tucson, AZ  85721  USA\\
     $^4$ Department of Physics, University of Hawaii, Honolulu, HI 96822  USA}

\begin{abstract}
Dynamical dark matter (DDM) is a new framework for dark-matter physics in 
which the dark sector comprises an ensemble of individual component fields 
which collectively conspire to act in ways that transcend those normally 
associated with dark matter.  Because of its 
non-trivial structure, this DDM ensemble --- unlike most 
traditional dark-matter candidates --- cannot be characterized in terms of 
a single mass, decay width, or set of scattering cross-sections, but 
must instead be described by parameters which describe the collective 
behavior of its constituents.  Likewise, the components of such an ensemble 
need not be stable so long as lifetimes are balanced against cosmological 
abundances across the ensemble as a whole.
In this paper, we investigate the prospects for identifying a DDM ensemble
at the LHC and for distinguishing such a dark-matter candidate from the 
candidates characteristic of traditional dark-matter models.
In particular, we focus on DDM scenarios in which the component fields of the 
ensemble are produced at colliders alongside some number of Standard-Model 
particles via the decays of additional heavy fields.  The invariant-mass 
distributions of these Standard-Model particles turn out to possess several
unique features that cannot be replicated in most traditional dark-matter 
models.  We demonstrate that in many situations it is possible to differentiate 
between a DDM ensemble and a traditional dark-matter candidate 
on the basis of such distributions.    
Moreover, many of our results also apply more generally to a variety of 
other extensions of the Standard Model which involve multiple stable or 
metastable neutral particles.
\end{abstract}

\maketitle

\newcommand{\newc}{\newcommand}
\newc{\gsim}{\lower.7ex\hbox{$\;\stackrel{\textstyle>}{\sim}\;$}}
\newc{\lsim}{\lower.7ex\hbox{$\;\stackrel{\textstyle<}{\sim}\;$}}
\makeatletter
\newcommand{\biggg}{\bBigg@{3}}
\newcommand{\Biggg}{\bBigg@{4}}
\makeatother

\def\vac#1{{\bf \{{#1}\}}}

\def\beq{\begin{equation}}
\def\eeq{\end{equation}}
\def\beqn{\begin{eqnarray}}
\def\eeqn{\end{eqnarray}}
\def\calM{{\cal M}}
\def\calV{{\cal V}}
\def\calF{{\cal F}}
\def\half{{\textstyle{1\over 2}}}
\def\quarter{{\textstyle{1\over 4}}}
\def\ie{{\it i.e.}\/}
\def\eg{{\it e.g.}\/}
\def\etc{{\it etc}.\/}


\def\inbar{\,\vrule height1.5ex width.4pt depth0pt}
\def\IR{\relax{\rm I\kern-.18em R}}
 \font\cmss=cmss10 \font\cmsss=cmss10 at 7pt
\def\IQ{\relax{\rm I\kern-.18em Q}}
\def\IZ{\relax\ifmmode\mathchoice
 {\hbox{\cmss Z\kern-.4em Z}}{\hbox{\cmss Z\kern-.4em Z}}
 {\lower.9pt\hbox{\cmsss Z\kern-.4em Z}}
 {\lower1.2pt\hbox{\cmsss Z\kern-.4em Z}}\else{\cmss Z\kern-.4em Z}\fi}
\def\TBBN{T_{\mathrm{BBN}}}
\def\OmegaCDM{\Omega_{\mathrm{CDM}}}
\def\OmegaDM{\Omega_{\mathrm{CDM}}}
\def\Omegatot{\Omega_{\mathrm{tot}}}
\def\rhocrit{\rho_{\mathrm{crit}}}
\def\tnow{t_{\mathrm{now}}}
\def\arcsinh{\mbox{arcsinh}}
\def\Omegatotnow{\Omega_{\mathrm{tot}}^\ast}
\def\mij{m_{jj}}
\def\mijmin{m_{jj}^{(\mathrm{min})}}
\def\epsig{\epsilon_{\mathrm{sig}}}
\def\Lint{\mathcal{L}_{\mathrm{int}}}
\newcommand{\Dsle}[1]{\hskip 0.09 cm \slash\hskip -0.23 cm #1}
\newcommand{\Dirsl}[1]{\hskip 0.09 cm \slash\hskip -0.20 cm #1}
\newcommand{\met}{{\Dsle E_T}}


\input epsf





\section{Introduction\label{sec:intro}}


Recently, a new framework for dark-matter physics has been 
proposed~\cite{DynamicalDM1,DynamicalDM2}.  This new framework is called 
``dynamical dark matter'' (DDM), and previous discussions of DDM have 
focused on its overall theoretical properties~\cite{DynamicalDM1} as well 
as on the theoretical and phenomenological implications of a specific 
model~\cite{DynamicalDM2,DynamicalDM3} that was constructed within this 
new framework.  In this paper, we shall return to consideration of the 
general DDM framework as a whole, and discuss the detection 
prospects for a broad class of DDM models at colliders, and in 
particular at the Large Hadron Collider (LHC).

As discussed in Refs.~\cite{DynamicalDM1,DynamicalDM2}, the central 
hallmark of the dynamical dark-matter framework is that the dark sector 
consists not of one (or merely a few) stable dark-matter particles, but rather an 
{\it ensemble}\/ of constituents which act collectively in ways that 
transcend the physics normally associated with more traditional dark 
sectors.  Such a DDM ensemble is not randomly assembled, but instead has 
certain internal structures which guarantee its phenomenological 
viability.  For example, stability on cosmological time scales --- 
normally considered to be a sacrosanct property for traditional 
dark-matter candidates --- is not a requirement for such an ensemble.  
Instead, lifetimes are balanced against cosmological abundances across the 
different constituents of this ensemble in such a way that those 
components with larger decay widths (and consequently shorter lifetimes) 
necessarily have smaller cosmological abundances, and vice versa.  
Indeed, as discussed in Refs.~\cite{DynamicalDM2,DynamicalDM3}, this 
balancing represents a novel way of satisfying phenomenological bounds 
on the dark sector without imposing stability as a whole, and ultimately 
represents the most general dark sector that can be imagined.

Most traditional dark-matter candidates can be characterized in terms of 
their masses and couplings to Standard-Model (SM) 
states, and indeed most phenomenological bounds in the dark-matter literature 
are phrased in terms of constraints on these variables~\cite{reviews}.
By contrast, in the DDM framework the dark-matter ``candidate'' is the 
entire ensemble, and the parameters which ultimately characterize a DDM ensemble
describe not only the couplings of its individual constituents to SM states, but 
also the internal structure of the ensemble itself.  In general, such an internal 
structure might consist of relationships between  
the masses, relic abundances, and couplings of its components.  As a result,
the natural parameters which characterize DDM models and their 
phenomenology are fundamentally different from those which suffice to  
describe traditional dark-matter candidates.  Indeed, the most 
fundamental parameters which characterize a DDM ensemble are those which
describe how quantities such as the constituent-particle masses, abundances, 
decay widths, and cross-sections scale with respect to one another 
across the ensemble as a whole.  This clearly represents
a new way of thinking about a dark sector, but
the need for such an approach is one of the primary features of the DDM framework.

In this paper, we shall focus on the implications of these fundamental 
differences for the collider phenomenology of DDM ensembles, and in particular 
for their discovery potential at the LHC.
The canonical channels in which one generally expects to obtain
evidence for a DDM ensemble involve substantial missing transverse 
energy (hereafter denoted $\met$) --- just as is the case for a traditional 
dark-matter candidate.  It is therefore crucial to develop 
strategies for distinguishing between these two classes of models once an 
excess in one or more of these $\met$ channels has been identified.  
One such strategy, which was discussed in 
Ref.~\cite{DynamicalDM3}, is to search for correlations between 
$\met$ signatures and signals in other channels to which 
the DDM ensemble might simultaneously give rise --- channels 
not normally associated with dark matter.  Indeed, in DDM models, only 
those portions of the DDM ensemble which are stable on collider time scales 
contribute to missing-energy signals at colliders.  By contrast, it is possible 
that other portions of the DDM ensemble will have much shorter lifetimes and 
therefore manifest themselves in different channels entirely --- channels that
may not involve $\met$ whatsoever.  As discussed in Ref.~\cite{DynamicalDM3}, a wide 
variety of DDM scenarios generically give rise to observable excesses in both 
classes of channels simultaneously.  However, in a variety of other DDM contexts, 
such correlations among channels may not be possible due to the particulars of 
the model and the lifetimes of the constituent fields in the ensemble.  In such 
contexts --- and especially when those constituent fields 
manifest themselves only through $\met$ signatures --- alternative strategies 
for identifying DDM ensembles are necessary.

Such alternative strategies will be the primary focus of this paper.  In particular, 
one of our central aims is to demonstrate that in many cases it is 
possible to distinguish DDM ensembles from more traditional dark-matter 
candidates based solely on results from channels in which the dark-matter particles 
manifest themselves as $\met$ alone.  
For example, the invariant-mass distributions of SM states produced in association with
the constituent fields in a DDM ensemble by the decays of heavy particles 
can exhibit qualitative features which transcend those usually associated with 
traditional dark-matter candidates.  As we shall see, such invariant-mass distributions 
can therefore provide a powerful experimental discriminant between DDM ensembles 
and these traditional candidates.  Furthermore, because this technique relies solely on
signatures in detection channels in which the component fields of the DDM ensemble appear 
as $\met$, these signatures are insensitive to the 
precise lifetimes of any component fields which leave an imprint 
on the invariant-mass distribution in question.   Hence, they are likewise 
insensitive to the characteristic instability of the DDM ensemble as a whole, and
this remains true provided that these fields are all sufficiently 
long-lived so as not to decay within the detector volume.  
These aspects of our discussion therefore have a broad applicability even beyond 
the context of the DDM framework, and apply quite 
generally to any multi-component dark-matter scenario involving an additional heavy 
``parent'' particle which decays to final states involving the dark-sector fields, 
or to any scenario involving multiple metastable neutral fields and such a parent
particle.

In order to explicitly illustrate these themes, in this paper we shall consider the case 
in which each parent particle decays directly to a single constituent field within 
the DDM ensemble along with a pair of strongly-interacting SM particles (\ie, quarks or gluons).  
Even in this simplest non-trivial case, we shall demonstrate that there exist a range
of characteristic features which are imprinted on the invariant-mass distribution of the 
two resulting jets and which can permit one to distinguish a DDM ensemble from any 
traditional dark-matter candidate.  Moreover, we shall show that there exist production
mechanisms for the parent particles (such as pair-production via strong interactions) 
with event rates sufficient to enable such a differentiation at the 
$5\sigma$ significance level within the first $30\mathrm{~fb}^{-1}$ of integrated 
luminosity at the $\sqrt{s}= 14$~TeV LHC.

This paper is organized as follows.  In Sect.~\ref{sec:production}, we 
outline the general aspects of the collider phenomenology of DDM 
ensembles and discuss the strategies for indirectly observing those 
ensembles in different classes of DDM models from the perspective of
effective-operator analysis.  In doing so, we devote particular attention
to models in which the component particles in the DDM ensemble
are produced via the decays of a heavy parent particle.  In 
Sect.~\ref{sec:mijDistributions}, we calculate the invariant-mass 
distributions associated with pairs of SM states produced in conjunction with 
each dark-matter particle via this mechanism and compare them to the invariant-mass 
distributions obtained in theories involving only a single particle stable 
on collider time scales.  In Sect.~\ref{sec:significances}, we assess the 
statistical significance with which such non-traditional invariant-mass 
distributions can be distinguished from those in more traditional dark-matter 
models.  In Sect~\ref{sec:productionchannels},
we provide an example of one production mechanism which naturally provides
event rates of the order required for such a differentiation within the first
$30\mathrm{~fb}^{-1}$ of integrated luminosity at the $\sqrt{s}=14$~TeV 
LHC --- namely, the pair-production of strongly-interacting parent 
particles with masses near the TeV scale.  Finally, in Sect~\ref{sec:conclusions}, 
we provide an assessment of how various subtleties associated with certain 
production mechanisms for the parent particle (such as the combinatorial 
background associated with incorrect pairings of final-state jets for processes 
which yield more than one such parent particle per event) are expected to impact 
our results.  We also comment on the broader applicability of our results to 
scenarios outside the DDM framework.


\section{Dynamical Dark Matter at the LHC: ~General Considerations\label{sec:production}}


One of the hallmarks of the DDM framework is that the dark sector consists of 
an ensemble of particles $\chi_n$, where $n=\{0,\ldots,N\}$, with $N$ presumed 
to be relatively large, \ie, $N\gg 1$.  For convenience, we shall label these
particles in order of increasing mass, \ie, $m_{n+1} \geq m_n$.
Moreover, as discussed in 
Refs.~\cite{DynamicalDM1,DynamicalDM2,DynamicalDM3}, these particles $\chi_n$   
exhibit a broad spectrum of decay widths $\Gamma_n$ which scale inversely 
with their corresponding cosmological abundances $\Omega_n$.
As a result of these different decay widths, different $\chi_n$ may manifest 
themselves in qualitatively different ways at colliders --- even in situations 
in which the $\chi_n$ all have similar quantum numbers and are therefore  
produced via similar processes.  Those $\chi_n$ with lifetimes 
$\tau_n \gtrsim 10^{-10}$~s are stable on collider time scales and appear 
in a collider detector as $\met$.  
By contrast, any of the $\chi_n$ with lifetimes $\tau_n \lesssim 10^{-10}$~s
decay within the detector volume.  Provided these rapidly-decaying 
$\chi_n$ decay predominately to final states involving SM particles, evidence 
for a DDM ensemble could potentially be obtained via the observation of signals 
in complementary channels which individually provide evidence for $\chi_n$ within either
of these $\tau_n$ regimes.  Such signatures are discussed in 
Ref.~\cite{DynamicalDM3}.  However, for models in which $\tau_n \gtrsim 10^{-10}$~s for 
all $\chi_n$ --- or for models in which the $\chi_n$ that decay within the detector
volume are not collectively produced at rates sufficient to yield observable effects at 
colliders --- such multi-channel correlation techniques are not useful.  In such cases, 
alternative strategies must be found for distinguishing DDM models from more 
traditional dark-matter models. 
         
In this paper, we will examine one such strategy, which is applicable to a 
broad class of DDM models possessing two key characteristics.  First, in addition 
to the constituent fields $\chi_n$ of the DDM ensemble, the field content of the 
model must include one or more heavy particles $\psi$ which can be 
produced at a substantial rate at a hadron collider.   
For example, if the $\psi$ transform non-trivially under the SM $SU(3)_c$ 
gauge group, they can be produced copiously via their interactions 
with the quark and gluon fields.  Second, these additional, heavy particles 
must decay with a sizeable branching fraction into final states 
including at least two SM fields, along with one or more of the $\chi_n$.
Decay topologies of this sort arise generically, for example, in a specific class of
DDM scenarios in which both the parent particle $\psi$ and the constituents of  
the DDM ensemble are charged under an approximate symmetry.  However, in such 
cases it is an important property of the general DDM framework that such a 
symmetry is neither required nor need be preserved exactly.  Therefore, the
ensemble constituents need not ultimately be stable. 

We shall demonstrate that it is possible to differentiate between 
traditional, single-particle models of dark matter and multi-component 
scenarios, such as those which arise in the DDM framework, by examining the 
invariant-mass distributions of the SM fields produced by $\psi$ decays.  
Of course, in cases in which only a small number of the $\chi_n$ are kinematically
accessible in $\psi$ decays, these invariant-mass distributions 
are distinguished by the presence of multiple kinematic edges.  Note that 
similar features arise in other contexts as well, most notably that in which
a parent particle can decay into final states involving different 
{\it multiplicities} of the same stable dark-matter 
particle~\cite{Kaustubh1,Kaustubh2}.  By contrast, in cases in which the 
number of kinematically-accessible $\chi_n$ is large and the decay 
phenomenology of the $\psi$ particles depends more sensitively on the full    
structure of the DDM ensemble, qualitatively different features 
emerge.  In particular, while individual kinematic edges are no longer manifest,
the invariant-mass distributions can exhibit distinctive shapes not realized 
in single-particle dark-matter scenarios.  These distributions can therefore     
provide a powerful experimental discriminant between DDM ensembles and  
more traditional dark-matter candidates.   

We note that the technique described above has a broad range of applicability
because it is not predicated on the observation of signals of
both collider-stable and promptly-decaying $\chi_n$, but rather of 
collider-stable fields alone.     
Moreover, for the same reason, this technique can also be 
applied more broadly to a wide variety of multi-component dark-matter models, 
or to other scenarios which involve multiple particles which are 
stable on collider time scales.  Such situations can arise in
certain limits of traditional dark-matter scenarios in which the dark-matter 
candidate is stabilized by a parity symmetry such as $R$-parity in 
supersymmetric models or KK parity~\cite{KKParity} in higher-dimensional 
theories in which the SM propagates in the 
bulk~\cite{Antoniadis,DDGLargeED,UED}.  For example, while only the lightest 
parity-odd particle is absolutely stable in such scenarios, situations can arise 
in which the decay rates of heavier parity-odd particles with similar 
quantum numbers are suppressed 
(either by kinematics or by some additional consideration) to such an extent
that they are also stable on collider time scales.  In such cases,
the invariant-mass distributions associated with the decays of even heavier 
fields in the theory include contributions from final states involving {\it all}\/
such stable or metastable particles.  These distributions can therefore yield 
valuable information about the overall coupling structure and mass spectrum of 
the theory.
      
In order to analyze the collider phenomenology of DDM ensembles, 
it is first necessary to characterize the properties of such 
dark-matter candidates in a straightforward and physically 
meaningful manner.  Indeed, a DDM ensemble is not a single particle or a small 
group of particles, but rather a vast collection of individual states whose
collective properties dictate the dark-matter phenomenology of the model.
Taken together, therefore, an ensemble of such states constitutes 
a dark-matter candidate which cannot be characterized in terms of a single 
well-defined mass, decay width, or set of cross-sections for processes involving 
SM fields.  Instead, a DDM ensemble is more aptly characterized by parameters
which describe its aggregate internal structure.  Such parameters may include, 
for example, scaling exponents in certain relations between masses and 
cross-sections, or between cosmological abundances and decay widths, that 
hold across the DDM ensemble as a whole.  More specifically, one natural set 
of such parameters might include the density of states expressed as a function of the 
masses $m_n$ of the $\chi_n$, and a set of exponents which describe the 
scaling behavior of the couplings of the $\chi_n$ to other fields present 
in the theory.  

In any arbitrary DDM model, such couplings can be described by a set of operators
$\mathcal{O}_{n_1,n_2,\ldots}^{(\alpha)}$, where the indices $n_i$ indicate the 
dark-sector fields $\chi_{n_i}$ involved and where the index $\alpha$ labels the 
operator in question.  We shall let $d_\alpha$ denote the mass dimension of the 
operator $\mathcal{O}_{n_1,n_2,\ldots}^{(\alpha)}$. 
We shall assume for the sake of simplicity in what follows that the only 
$\mathcal{O}_{n_1,n_2,\ldots}^{(\alpha)}$ which play a meaningful role in 
the collider phenomenology of the DDM ensemble are members of the subset
$\mathcal{O}_{n}^{(\alpha)}$ of operators which involve only a 
single $\chi_n$.  We therefore have an effective interaction Lagrangian of the
form 
\begin{equation}
    \mathcal{L}_{\mathrm{eff}} ~=~ \sum_{\alpha}\sum_{n=0}^{N}
    \frac{c_{n\alpha}}{\Lambda^{d_\alpha-4}}\mathcal{O}_{n}^{(\alpha)}+\ldots~,
    \label{eq:EffLagrangian}
\end{equation}
where $\Lambda$ is the cutoff scale of the effective theory, and
where $c_{n\alpha}$ is the dimensionless operator coefficient
associated with $\mathcal{O}_{n}^{(\alpha)}$.
In addition, we shall also assume that all the constituent 
particles in the ensemble have the same quantum numbers, and thus that the set of  
$\mathcal{O}_{n}^{(\alpha)}$ consistent with the symmetries of the theory 
shares a common operator structure for all $n$.   In other words, for any given 
$\alpha$, the only $n$-dependence appears in the mass $m_n$ of the field $\chi_n$ 
and the coefficient $c_{n\alpha}$.
Furthermore, we assume that the distribution of $c_{n\alpha}$ across the ensemble 
depends solely on $m_n$, and that both this distribution and the mass spectrum of the
ensemble exhibit general scaling relations of the form
\begin{eqnarray}
  m_{n} &=& m_0 + n^\delta \Delta m \nonumber \\
  c_{n \alpha} &=& c_{0 \alpha} \left(\frac{m_n}{m_0}\right)^{\gamma_\alpha}~,
  \label{eq:MassAndCouplingScalings}
\end{eqnarray}  
where the (positive) mass-splitting parameter $\Delta m$ and the scaling exponents 
$\delta$ and $\gamma_\alpha$ are free parameters.  Note that $\delta > 0$ by construction. 
All of these assumptions can, of course, be relaxed; however, 
the qualitative results obtained below still continue to hold in more general cases 
as well.  Note that the $\mathcal{O}_{n}^{(\alpha)}$ are ultimately responsible both for the 
production of the $\chi_n$ at colliders, and for their decay phenomenology --- both in 
terms of their lifetimes $\tau_n$ and in terms of their decay products.  Of 
course the $\mathcal{O}_{n}^{(\alpha)}$ which play a dominant role in the production of 
the $\chi_n$ need not be the same as those which play a dominant role in determining 
their decay properties.  

In this paper we shall focus primarily on the $\chi_n$ 
which are stable on collider time scales, as discussed above; hence our primary concern 
in what follows will be with production rather than decay.  In fact, given the 
scaling relations and parameterizations in Eqs.~(\ref{eq:EffLagrangian}) 
and~(\ref{eq:MassAndCouplingScalings}), we can now characterize precisely which 
components of a given DDM ensemble contribute to $\met$ signals once produced.
Assuming that a given $\chi_n$ decays primarily to final states comprising 
SM fields the sum of whose masses is much smaller than $m_n$, we find that 
$\Gamma_n$ scales roughly like
\begin{equation}
  \Gamma_{n} ~\sim~ \sum_{\alpha}
  c_{n \alpha}^2\frac{m_n^{2d_\alpha-7}}{\Lambda^{2d_\alpha-8}}~.
  \label{eq:DecayRateSchematic}
\end{equation}
As discussed above, the distribution of $\Gamma_n$ across the
DDM ensemble has a significant impact not only on the cosmological aspects of 
that ensemble, but on its collider phenomenology as well. 
When the lifetime $\tau_n = 1/\Gamma_n$ of the particle is short on collider time scales 
($\tau_n \lesssim 10^{-12}$~s), the particle decays promptly within the detector.  When 
the lifetime is long ($\tau_n \gtrsim 10^{-10}$~s), the particle decays outside the 
detector and appears as $\met$.  
In the intermediate region 
($10^{-12}\mathrm{~s}\lesssim \tau_n \lesssim 10^{-10}\mathrm{~s}$), the particle may 
give rise to a displaced vertex.  Thus $\chi_n$ within different ranges of $m_n$ in the 
ensemble may manifest themselves either as $\met$ or via their decay 
products, depending on the value of $\Gamma_n$.  For example, in the case in which 
a single operator $\mathcal{O}_{n}^{(\alpha)}$ governs the decay width of all relevant 
$\chi_n$, Eqs.~(\ref{eq:MassAndCouplingScalings}) and~(\ref{eq:DecayRateSchematic}) 
together imply that the requirement for appearing as $\met$ is          
\begin{equation}
  \Gamma_0 \left(1+n^\delta\frac{\Delta m}{m_0}\right)^{2d_\alpha-7+2\gamma_\alpha}
  ~\lesssim~ 6.58 \times 10^{-15}\mathrm{~GeV} ~,
  \label{eq:ChinLifetimeReqment}
\end{equation}
where $\gamma_\alpha$ here denotes the particular scaling exponent associated 
with the operator which effectively controls the decay width of $\chi_n$, and 
$\Gamma_0$ is the decay width of the lightest state in the ensemble.  


\section{Imprints of DDM Ensembles in Kinematic Distributions\label{sec:mijDistributions}}


By and large, when a dark-sector field is produced at the LHC via the
decay of a heavy parent particle $\psi$, multiple SM fields can also be 
produced via the same decay.
One useful variable that can assist in distinguishing models involving
a single dark-matter candidate from those involving more complicated dark sectors
in scenarios of this sort is the invariant mass of the additional SM particles 
produced by decays of the parent particle.
The identification of features in invariant-mass distributions has 
been shown to be effective in differentiating between the different symmetries
which might stabilize a dark-matter particle in traditional 
dark-matter models~\cite{Kaustubh1}.  In this paper, we show that it is also 
effective in distinguishing between traditional and DDM dark sectors.
  
For purposes of illustration, we focus on the case in which each $\psi$
decay yields two SM particles and a single $\chi_n$.  Indeed, this is the 
simplest case in which a non-trivial invariant-mass distribution is 
obtained for the SM particles from the decay of a single $\psi$.
Furthermore, since strongly-interacting particles can be produced copiously at 
colliders, we shall focus on the case in which $\psi$ carries $SU(3)_c$ charge and 
decays to a pair of SM quarks or gluons which form hadronic jets, and 
the relevant kinematic variable is the invariant mass $\mij$ of the two jets 
thus produced.  We assume that each $\psi$ decays primarily via 
three-body processes of the form $\psi \rightarrow j j\chi_n$.  
We discuss the alternative possibility in which $\psi$ decays to 
the same final state through cascade decays involving a on-shell 
intermediary in Sect.~\ref{sec:conclusions}.  Finally, we assume here 
for simplicity that the pair of final-state jets produced via the decay 
of each particular $\psi$ can be correctly identified in each event.
Such an identification is trivial in situations such as those in which $\psi$ is 
produced singly or in which there exist other decay channels for $\psi$ with branching 
fractions similar to $\sum_n\mathrm{BR}(\psi\rightarrow  j j\chi_n)$ whose 
decay products are readily distinguishable from jets.
However, if no such alternative decay channels exist --- \ie, if decays of the form 
$\psi\rightarrow  j j \chi_n$ dominate the width of $\psi$ --- the analysis becomes 
significantly more complex because of the non-trivial combinatorial 
issues which arise due to the possibility of incorrect pairings among the
final-state jets.  However, a number of techniques have been developed 
which can assist in identifying the correct jet pairings in such situations.  We 
shall discuss these techniques, and the effect of combinatorial issues in general, 
in more detail in Sect.~\ref{sec:conclusions}.  

We begin our analysis of the invariant-mass distributions which arise in the 
context of the DDM framework by briefly reviewing the characteristics of the 
corresponding distributions which arise in traditional dark-matter models.
This will be important for purposes of comparison when we consider the contrasting 
case of a full DDM ensemble.  Towards this end, 
let us consider a traditional dark-matter model in which a parent particle 
$\psi$ likewise decays predominately into a three-body final state comprising a dark-matter 
particle $\chi$ and a pair of strongly-interacting SM fields (either quarks or gluons).
The latter appear in the detector as hadronic jets, here labeled $j_1$ and $j_2$.
Irrespective of the Lorentz or $SU(3)_c$ representations of these fields,     
the differential partial width $d\Gamma_\psi \equiv d\Gamma(\psi\rightarrow j_1 j_2\chi)$ 
associated with this decay channel is given by the general three-body decay-width 
formula
\begin{equation}
  d\Gamma_\psi ~=~ \frac{1}{32 (2\pi)^3 m_\psi^3}
     \overline{|\mathcal{M}|}^2 d\mij^2 dm_{j_2\chi}^2~,
  \label{eq:dGammaEq}
\end{equation} 
where $\overline{|\mathcal{M}|}^2$ is the matrix element for the parton-level
process (averaged over the spin and color states of $\psi$), $\mij$ is the invariant 
mass of $j_1$ and $j_2$, and $m_{j_2\chi}$ is the invariant mass of $j_2$ and 
$\chi$.  The normalized invariant-mass distribution associated with these decays
is therefore given by 
\begin{equation}
  \frac{1}{\Gamma_\psi }\frac{d\Gamma_\psi }{d\mij} ~=~ 
  \frac{\mij}{16 (2\pi)^3 m_\psi^3\Gamma_\psi}\int^{m^2_+}_{m^2_-}
     \overline{|\mathcal{M}|}^2 dm_{j_2\chi}^2~,
  \label{eq:InvtMassDistNormalized}
\end{equation}
where the limits of integration are 
\begin{equation}
   m^2_{\pm} ~\equiv~ \frac{1}{2}\bigg[
   m_\psi^2 + m_\chi^2-\mij^2 \pm \sqrt{m_\psi^4 - 
   2m_\psi^2(m_\chi^2 + \mij^2) +(\mij^2 - m_\chi^2)^2}\bigg] 
\end{equation}
in the limit that parton masses can be neglected.  Note that
Eq.~(\ref{eq:InvtMassDistNormalized}) vanishes for $\mij = 0$ and for
$\mij = m_\psi-m_\chi$. 

\begin{figure}[t!]
\begin{center}
  \epsfxsize 3.0 truein \epsfbox {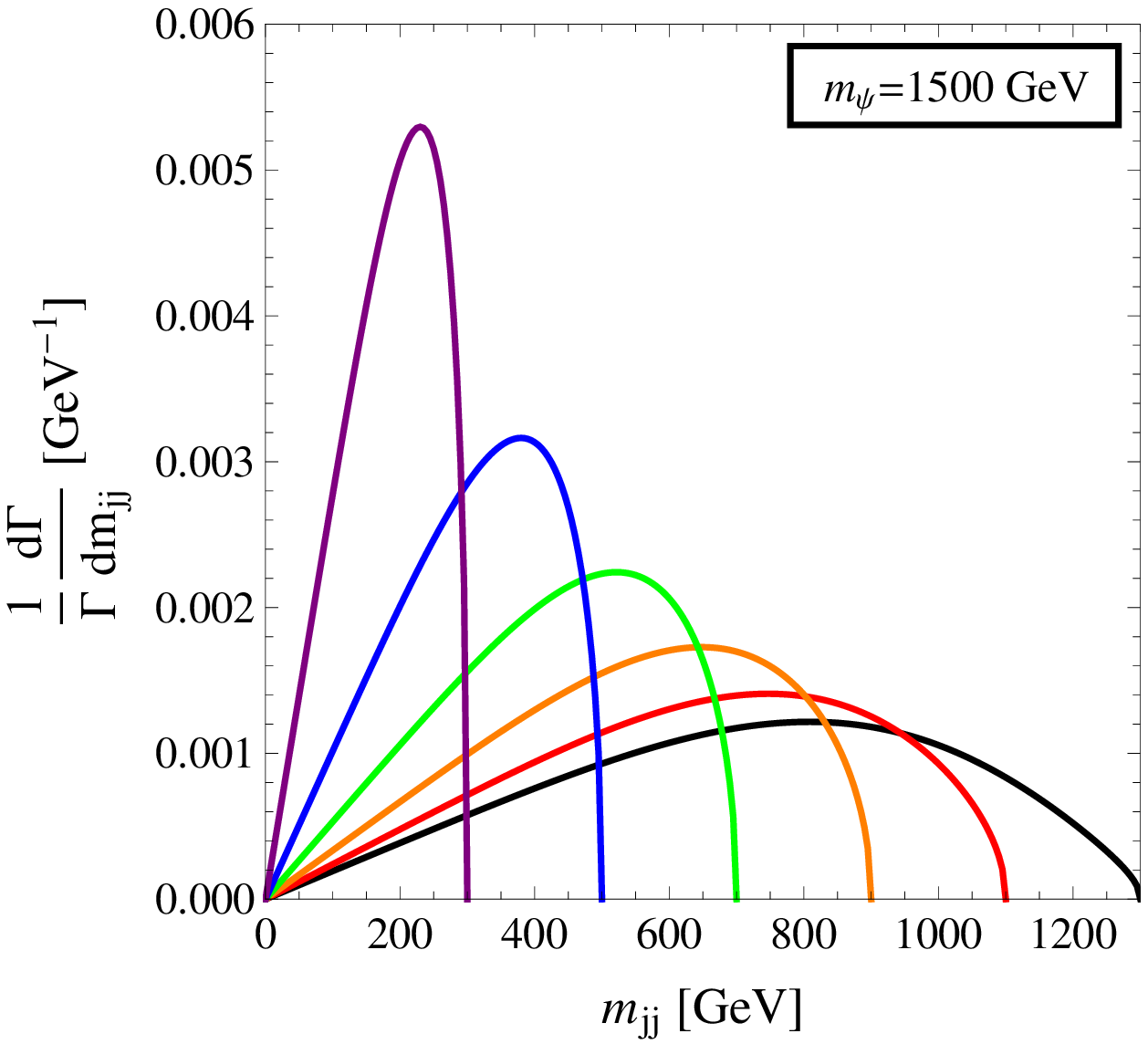}~~~~
  \epsfxsize 3.2 truein \epsfbox {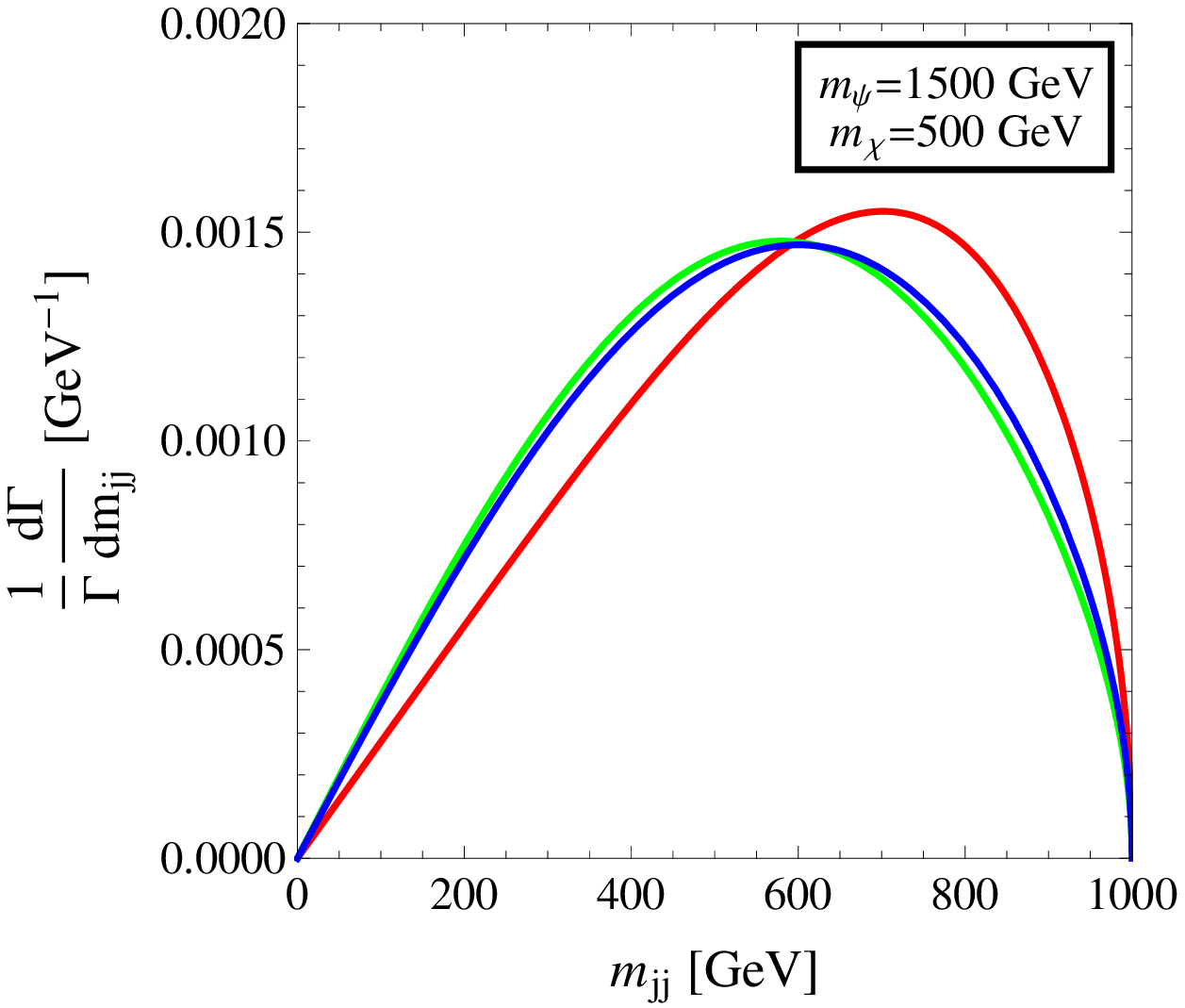}
\end{center}
\caption{Characteristic invariant-mass distributions in the traditional 
dark-matter models discussed in the text.  The left panel shows
the normalized dijet invariant-mass distributions associated with 
the decays of a heavy octet fermion $\psi$ with a mass $m_\psi = 1500$~GeV 
to a quark-antiquark pair and a single dark-matter candidate with a mass $m_\chi$.
The black, red, orange, green, blue, and purple curves correspond respectively to
$m_\chi = \{200,400,600,800,1000,1200\}$~GeV.  The right panel shows
the distributions for fixed $m_\psi = 1500$~GeV and $m_\chi = 500$~GeV, but for 
$\psi$ and $\chi$ with different spins and transformation properties under 
$SU(3)_c$ and therefore different coupling structures with the SM fields.  
The red curve corresponds to the parton-level process 
$\psi\rightarrow q\bar{q}\chi$ in which $\psi$ and $\chi$ are both fermions, the
green curve corresponds to the process $\psi\rightarrow qq\chi$ in which 
$\psi$ and $\chi$ are both scalars, and the blue curve corresponds to the 
process $\psi\rightarrow gq\chi$ in which $\psi$ is a fermion and $\chi$ is 
a scalar.  We observe that coupling structure does not induce a dramatic change in
the dijet invariant-mass distribution.
\label{fig:StdDMmijCurves}}
\end{figure}

The explicit form of the $\mij$ distribution which arises in any particular model
depends on the coupling structure between $\psi$, $\chi$, and the SM quark or gluon 
fields.  As a concrete example, let us consider the case in which $\chi$ and $\psi$
are both fermions, and $\psi$ transforms in the octet representation of $SU(3)_c$,
while $\chi$ is neutral under all SM symmetries.
In this case, $\psi$ can decay via the effective four-fermion interaction term 
\begin{equation}
  \mathcal{L}_{\mathrm{eff}} ~=~ \frac{c}{\Lambda^2}
      (\overline{q}_it^a_{ij}\psi^a)(\overline{\chi} q_j) +\mathrm{h.c.}
     \label{eq:FourFermiOpStdDM}
\end{equation}
where $q$ denotes a SM quark, $t_{ij}^a$ is the generator of $SU(3)$ in the 
fundamental representation, the indices $i$ and $j$ label the states in the 
fundamental representation, and $a$ labels the states in the adjoint representation.
The spin- and color-averaged squared matrix element for an interaction of this form 
is then given by
\begin{equation}
   \overline{|\mathcal{M}|}^2 ~=~ \frac{c^2}{\Lambda^4}
      (m_\psi^2-m_{j_2 \chi }^2)(m_{j_2 \chi}^2-m_\chi^2)~.
    \label{eq:MatElSqdStdDM}
\end{equation} 
In the left panel of Fig.~\ref{fig:StdDMmijCurves}, we display the $\mij$ 
distributions obtained in this case for $m_\psi = 1500$~GeV and several
different choices of $m_\chi$.  Note that each distribution shown features 
a characteristic mass edge at the kinematic endpoint $\mij = m_\psi - m_\chi$,
which differs for each $m_\chi$.  However, in most other aspects, the 
distributions are qualitatively quite similar, and in particular have 
the same overall shape. 

Of course, different coupling structures from the one specified in 
Eq.~(\ref{eq:FourFermiOpStdDM}) correspond to different functional forms
for $\overline{|\mathcal{M}|}^2$, and one might wonder what effect such 
differences in coupling structure have on the $\mij$ distribution as a whole.
In the right panel of Fig.~\ref{fig:StdDMmijCurves}, 
we compare the $\mij$ distributions associated with several allowed 
combinations of Lorentz and $SU(3)_c$ representations for both parent and 
daughter particles, which result in different coupling structures.  
The distributions shown correspond to the parton-level 
process $\psi\rightarrow q\bar{q}\chi$ for which $\psi$ and $\chi$ are both fermions
(red), the process $\psi\rightarrow qq\chi$ for which $\psi$ and $\chi$ are 
both scalars (green), and the process $\psi\rightarrow gq\chi$ for which 
$\psi$ is a fermion and $\chi$ is a scalar (blue).  These decay processes 
arise in the cases in which $\psi$ transforms as an ${\bf 8}$, as a {\bf 6}, and 
as a ${\bf 15}$ under $SU(3)_c$, respectively.  In each case, we have
set $m_\psi = 1500$~GeV and $m_\chi = 500$~GeV.  While these 
distributions vary slightly from one another, we once again see that 
they all have essentially the same shape, which includes a characteristic 
mass edge at $\mij = m_\psi - m_\chi$.  Indeed, we observe that the $\mij$ 
distributions do not differ dramatically among single-particle dark-matter 
models as a result of differences in coupling structure.

We now consider how the decay phenomenology of $\psi$ in the context of the 
DDM framework differs from that associated with the traditional 
dark-matter scenarios discussed above.  As we shall see, while each individual 
constituent $\chi_n$ in the DDM ensemble contributes to the width of $\psi$ 
and to the overall $\mij$ distribution in a manner analogous to a traditional 
dark-matter candidate, the collective behavior of these $\chi_n$
gives rise to distinctly new phenomena.  For concreteness, we once again focus on 
the particular case in which $\psi$ is a color-octet fermion and the 
$\chi_n$ are fermions which transform as singlets under the SM gauge group;   
however, as the right panel of Fig.~\ref{fig:StdDMmijCurves} attests, the 
shape of the $\mij$ distributions does not depend sensitively on the spins of the
particles involved or the structure of the interaction vertex, and the results 
obtained for different cases will therefore be analogous.  We focus on the case in
which $\psi \rightarrow j j\chi_n$ decays arise predominately due to a four-fermion 
interaction analogous to Eq.~(\ref{eq:FourFermiOpStdDM}):
\begin{equation}
  \mathcal{L}_{\mathrm{eff}} ~=~ \sum_n\bigg[\frac{c_{n}}{\Lambda^2}
      (\overline{q}_it^a_{ij}\psi^a)(\overline{\chi}_n q_j) +\mathrm{h.c.}\bigg]~.
     \label{eq:FourFermiOp}
\end{equation} 
For convenience, we shall henceforth use the symbol $\gamma$ to denote the particular 
scaling exponent $\gamma_{\alpha}$ associated with this operator, so that the 
operator coefficients $c_n$ scale according to the relation  
\begin{equation}
  c_n = c_0 \left(\frac{m_n}{m_0}\right)^\gamma~,
  \label{eq:DefOfgammaIndex}
\end{equation}
in the manner described in Eq.~(\ref{eq:MassAndCouplingScalings}).
We will also assume for simplicity that there are no additional decay channels for 
$\psi$, and that the sum of the branching fractions for all decays 
of the form $\psi \rightarrow j j\chi_n$ is effectively unity.
Moreover, we will assume that the operators which contribute 
to the decay widths of all $\chi_n$ with masses $m_n < m_\psi$ are sufficiently 
suppressed so that all such particles amply satisfy the condition in
Eq.~(\ref{eq:ChinLifetimeReqment}) and therefore manifest themselves as $\met$.

For the coupling structure specified in Eq.~(\ref{eq:FourFermiOp}), 
the differential decay width 
$d\Gamma_{\psi n}\equiv\Gamma(\psi\rightarrow \overline{q}q\chi_n)$ for $\psi$ 
decay into any particular $\chi_n$ is likewise given by Eq.~(\ref{eq:dGammaEq}), and  
$\overline{|\mathcal{M}|}^2$ is given by 
\begin{equation}
   \overline{|\mathcal{M}|}^2 
      ~=~ \frac{c_n^2}{\Lambda^4}
      (m_\psi^2-m_{j_2 n }^2)(m_{j_2 n}^2-m_{n}^2)~,
\end{equation} 
where $m_{j_2 n }$ denotes the invariant mass of $\chi_n$ and $j_2$.
Integrating over $dm_{j_2 n}^2$ yields 
the differential {\it partial} decay width of $\psi$ with respect to $\mij$ for 
the particular decay $\psi\rightarrow j j \chi_n$:
\begin{equation}
  \frac{d\Gamma_{\psi n}}{d\mij} ~=~ 
     \frac{c_n^2 \mij
     \sqrt{m_\psi^4 - 2m_\psi^2(\mij^2+m_n^2) 
        + (\mij^2 - m_n^2)}}{96(2\pi)^3 m_\psi^3\Lambda^4}
     \bigg[m_\psi^4 + m_\psi^2(\mij^2-2m_n^2) - 2\mij^4
     + \mij^2 m_n^2 + \mij^4\bigg]~.
\end{equation}
The partial width for $\psi\rightarrow j j \chi_n$ is therefore
\begin{equation}
  \Gamma_{\psi n} ~=~ \frac{c_n^2}{384 (2\pi)^3\Lambda^4m_\psi^3}
   \Bigg[m_\psi^8 - 8m_\psi^2m_n^2(m_\psi^4-m_n^4) 
      + 12m_\psi^4m_n^4 \ln\left(\frac{m_n^2}{m_\psi^2}\right)\Bigg]~.
\end{equation}
Since the products of each such decay mode appear in a collider detector as 
a pair of jets plus $\met$, the {\it total} differential dijet invariant-mass 
distribution observed is a sum of the $\mij$ distributions for each channel, 
weighted by the corresponding decay branching fraction: 
\begin{equation}
   \frac{1}{\Gamma_\psi}\frac{d\Gamma_\psi}{d\mij} ~=~ 
      \sum_{n=0}^{n_{\mathrm{max}}}\left(\frac{1}{\Gamma_{\psi n}}
      \frac{d\Gamma_{\psi n}}{d\mij} \times\mathrm{BR}_{\psi n}\right)~,
   \label{eq:TotalDiffWidth}
\end{equation} 
where $\Gamma_\psi \equiv \sum_n\Gamma_{\psi n}$ is the total contribution
to the decay width of $\psi$ from processes of the form 
$\psi\rightarrow jj\chi_n$, where     
$\mathrm{BR}_{\psi n}\equiv \Gamma_{\psi n}/\Gamma_\psi$ denotes the branching
fraction for the particular decay $\psi\rightarrow jj\chi_n$,
where $n_{\mathrm{max}}$ is the value of $n$ corresponding to the heaviest $\chi_n$  
kinematically accessible in $\psi$ decay.  Note that 
this result also applies in the case in which other decay channels 
exist for $\psi$ with distinguishable final states.
The invariant-mass distribution in Eq.~(\ref{eq:TotalDiffWidth}) therefore
depends not only on the parameters $\Delta m$ and $\delta$ 
(which control the mass spectrum of the $\chi_n$ and therefore the 
available phase space for each individual decay channel), but also the
parameter $\gamma$ (which affects the branching fractions associated with
these channels). 
   
\begin{figure}[ht!]
\begin{center}
  \epsfxsize 2.25 truein \epsfbox {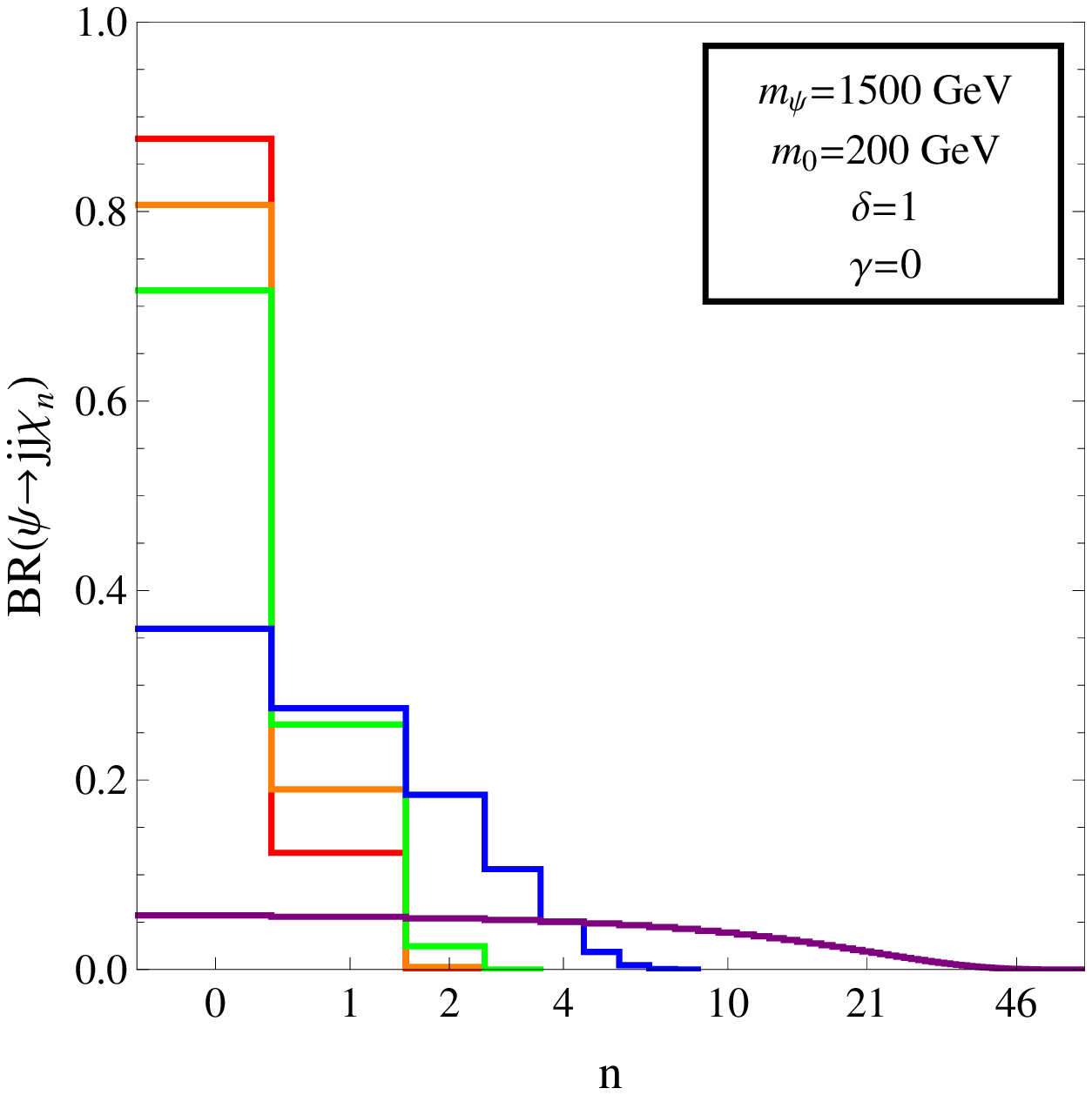}~~
  \epsfxsize 2.25 truein \epsfbox {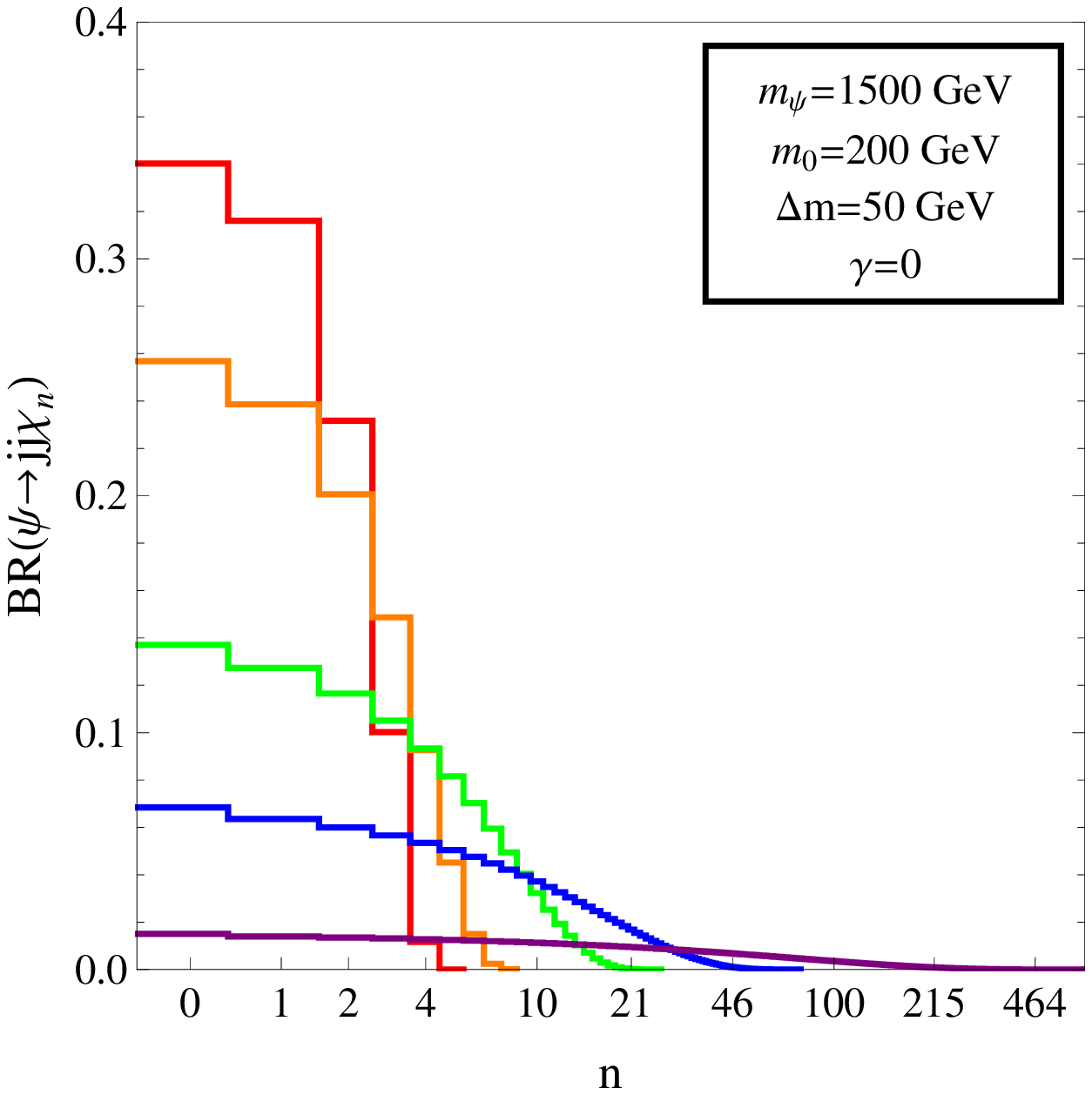}~~
  \epsfxsize 2.25 truein \epsfbox {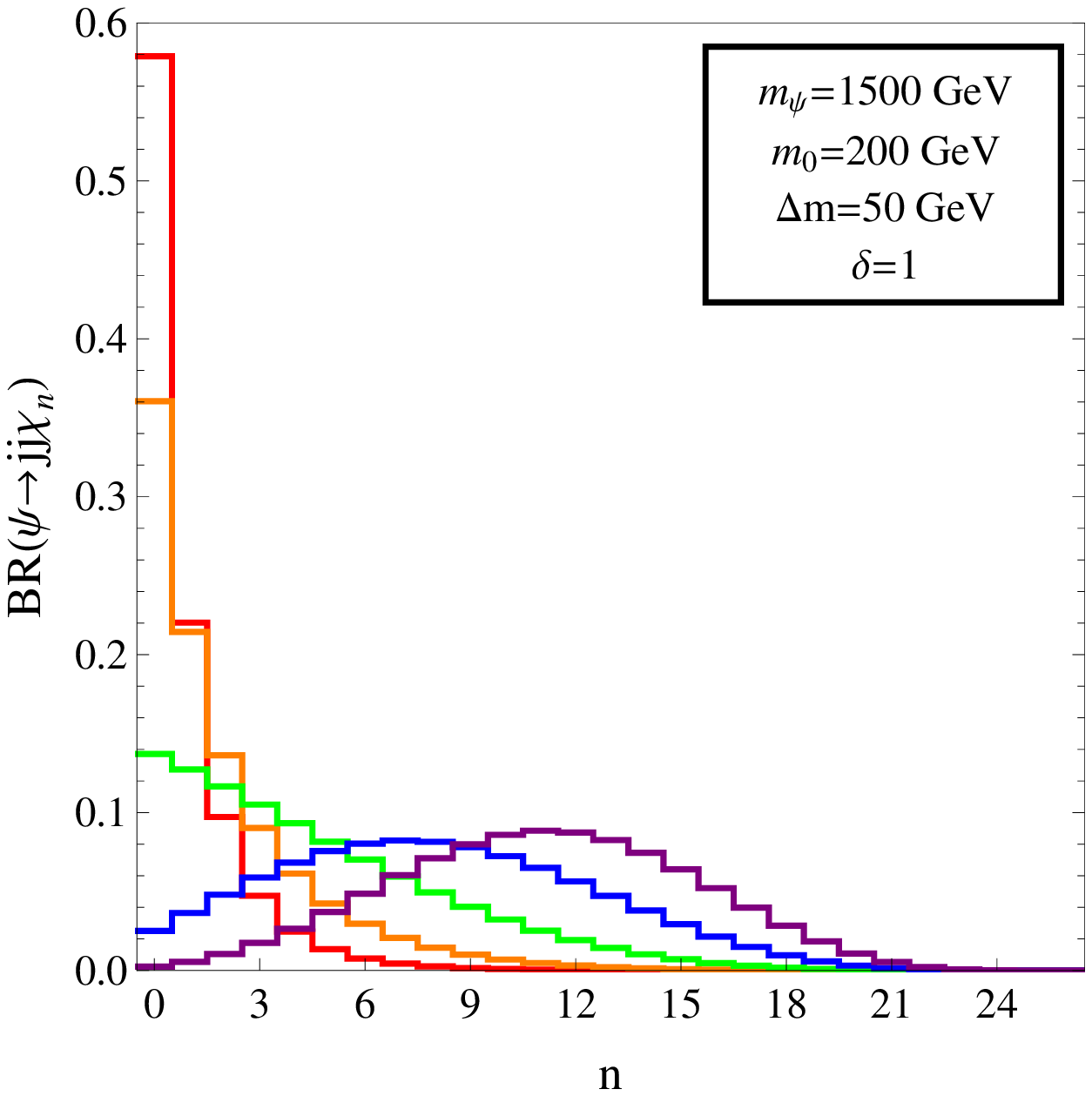}
\end{center}
\caption{The branching fraction 
$\mathrm{BR}(\psi\rightarrow j j\chi_n)$, plotted as a function of the DDM-ensemble
index $n$.  In all cases we have taken $m_\psi = 1500$~GeV and $m_0 = 200$~GeV.  In the
left panel, we have set $\delta =1$ and $\gamma =0$, and the red, orange, green, 
blue, and purple curves correspond respectively 
to $\Delta m = \{600,500,400,150,20\}$~GeV.  
In the center panel, we have set $\Delta m = 50$~GeV and $\gamma = 0$, and
the same colors respectively correspond to $\delta = \{2,1.5,1,0.75,0.5\}$.
In the right panel, we have set 
$\Delta m = 50$~GeV and $\delta = 1$, and the same colors respectively correspond
to $\gamma = \{-2,-1,0,1,2\}$.  Note that the left and center panels show the DDM 
index $n$ on a log scale, while the right panel shows $n$ on a linear scale.
\label{fig:BRDDM}}
\end{figure}

In order to explicitly illustrate the dependence of the branching 
fractions $\mathrm{BR}_{\psi n}$ on $\delta$, $\gamma$, and 
$\Delta m$, in Fig.~\ref{fig:BRDDM} we display $\mathrm{BR}_{\psi n}$ 
as a function of $n$.  
In each of the three panels shown, we have taken $m_\psi = 1500$~GeV and 
$m_0 = 200$~GeV.  The curves in the left panel correspond to different 
choices of $\Delta m$ for fixed $\gamma=0$ and $\delta=1$; this represents
a case in which the mass splitting between the $\chi_n$ is uniform and 
each of these particles couples to $\psi$ with equal strength.  In this case,  
the dependence of the branching fraction on $n$ is solely due to the 
the available phase space for $\psi \rightarrow jj\chi_n$ decays, which
decreases monotonically with $n$ up to the kinematic limit --- \ie, so long as 
$m_n \leq m_\psi$ --- above which $\mathrm{BR}_{\psi n} =0$.  For large 
values of $\Delta m$, we see that only a few states are kinematically 
accessible, and the branching fractions to the heavier states are 
significantly suppressed.  By contrast, for small $\Delta m$, a large number
of states are kinematically accessible, and $\mathrm{BR}_{\psi n}$ decreases 
quite gradually with increasing $n$.  A similar effect is manifest in the 
center panel of Fig.~\ref{fig:BRDDM}, in which the curves in which correspond to 
different values of $\delta$ for fixed $\gamma=0$ and $\Delta m =50$~GeV.
However, since increasing $\delta$ has the effect of increasing the mass 
gap between the $\chi_n$, we see that
$\mathrm{BR}_{\psi n}$ drops more rapidly with $n$ when $\delta$ is large.

Finally, in the right panel of Fig.~\ref{fig:BRDDM}, we display curves 
which correspond to different values of $\gamma$ for fixed $\delta=1$ and 
$\Delta m = 50$~GeV.  For $\gamma \leq 0$, we observe that    
$\mathrm{BR}_{\psi n}$ decreases monotonically with $n$ as in the left and
center panels, since such values of $\gamma$ simply imply an additional 
coupling suppression of $\mathrm{BR}_{\psi n}$ for the heavy states in 
addition to the phase-space suppression discussed above.  By contrast, 
the couplings of the heavy states are {\it enhanced} for $\gamma > 0$, and
the effect of this coupling enhancement and the phase-space suppression 
compete.  As a result, in this case, the dominant decay mode for $\psi$ 
may not be to the lightest state in the ensemble, but rather to a more 
massive state. 
 
\begin{figure}[t!]
\begin{center}
  \epsfxsize 2.25 truein \epsfbox {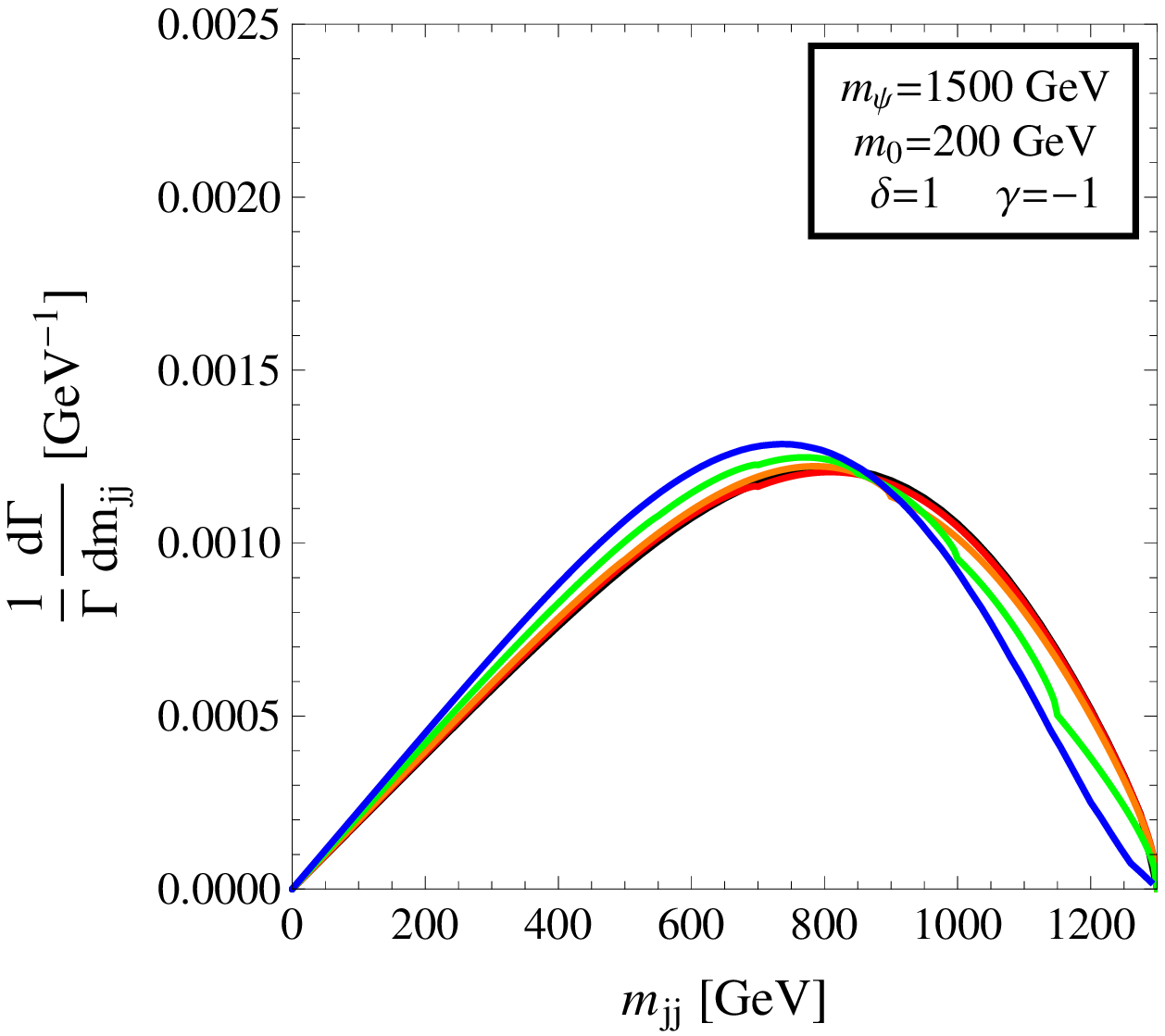}~~
  \epsfxsize 2.25 truein \epsfbox {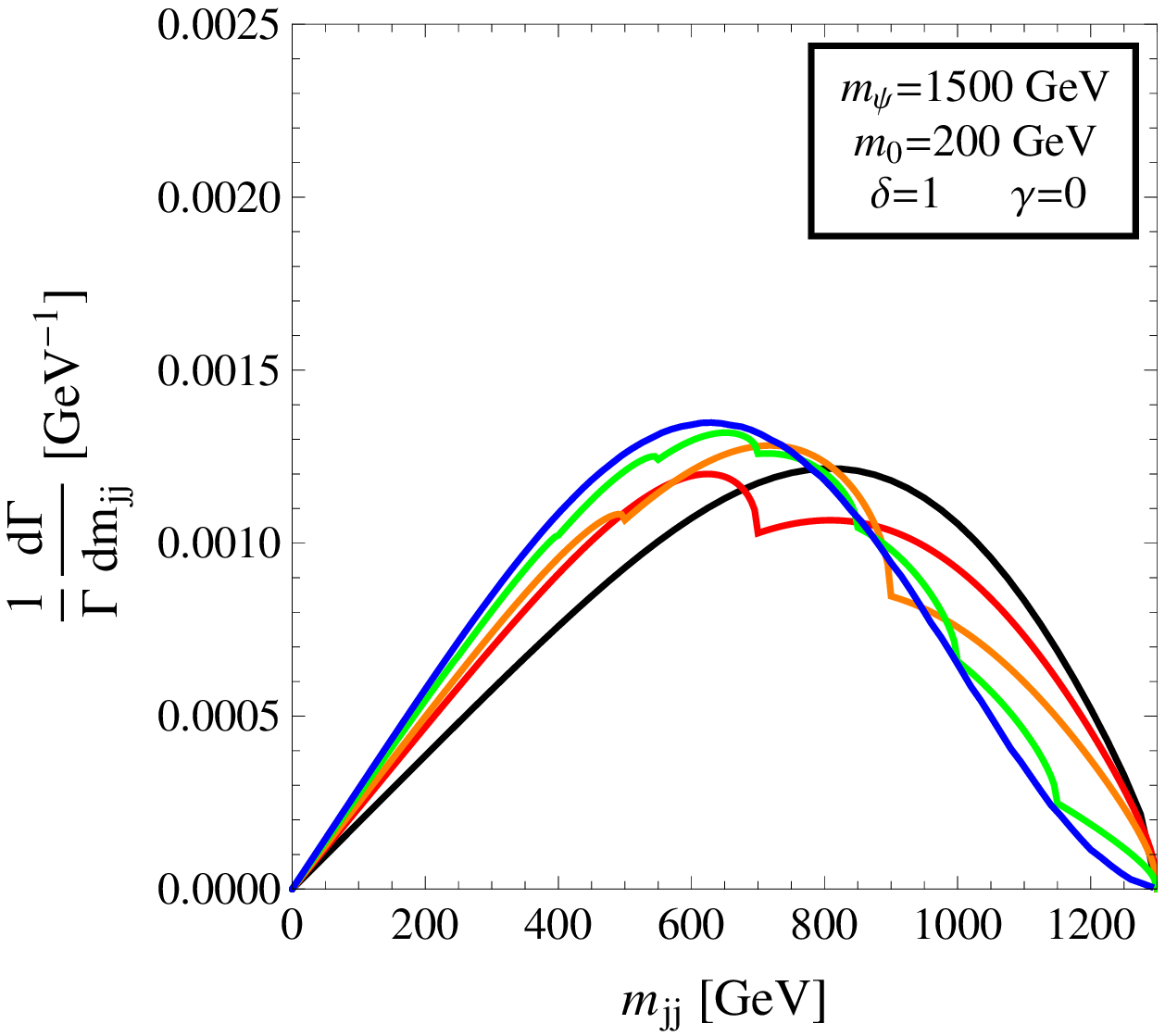}~~
\epsfxsize 2.25 truein \epsfbox {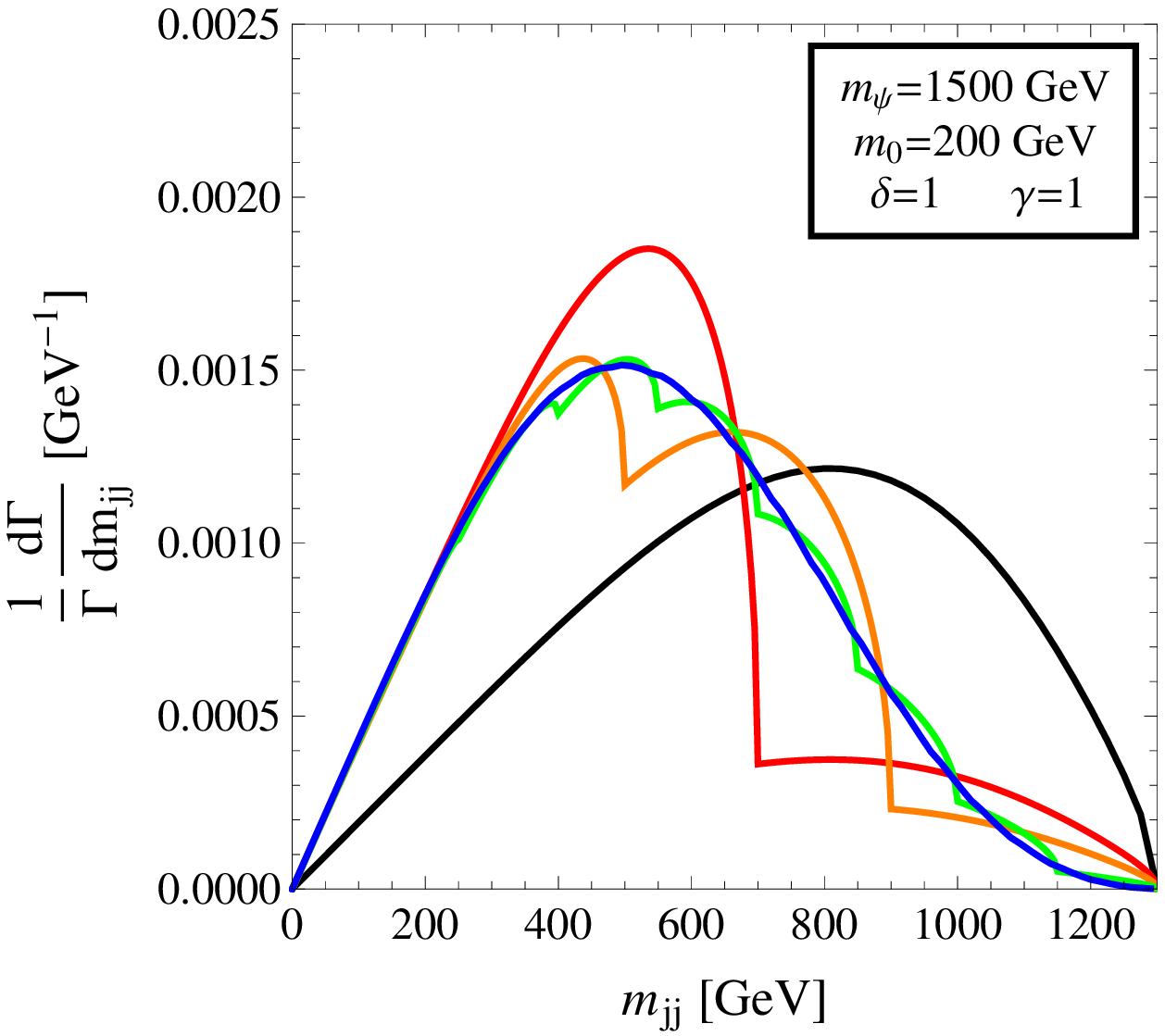}
\end{center}
\caption{DDM invariant-mass distributions, shown for increasing  
$\gamma$ with fixed $m_\psi = 1500$~GeV, $m_0=200$~GeV, and $\delta=1$.  The results
shown in the left, center, and right panels correspond to $\gamma=\{-1,0,1\}$, respectively.
In each panel, the red, orange, green, and blue curves correspond to 
mass splittings $\Delta m = \{600,400,150,20\}$~GeV, respectively, while  
the black curve shows the result for a traditional dark-matter candidate with 
$m_\chi = m_0$.  
\label{fig:mijDistsScaleGamma}}
\begin{center}
  \epsfxsize 2.25 truein \epsfbox {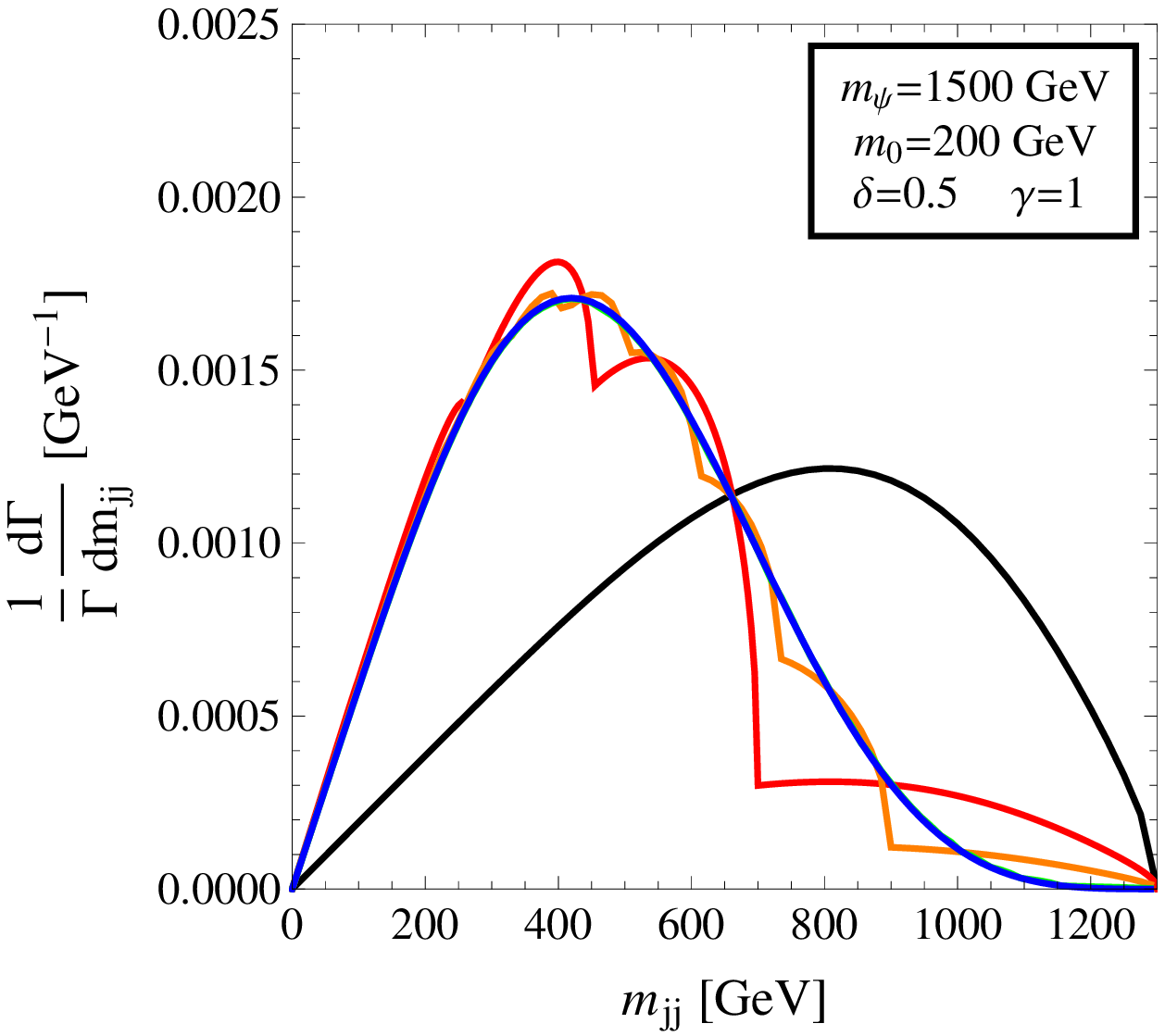}~~
  \epsfxsize 2.25 truein \epsfbox {mijDists_mpsi1500_m0x200_d1_g1_basic.eps}~~
  \epsfxsize 2.25 truein \epsfbox {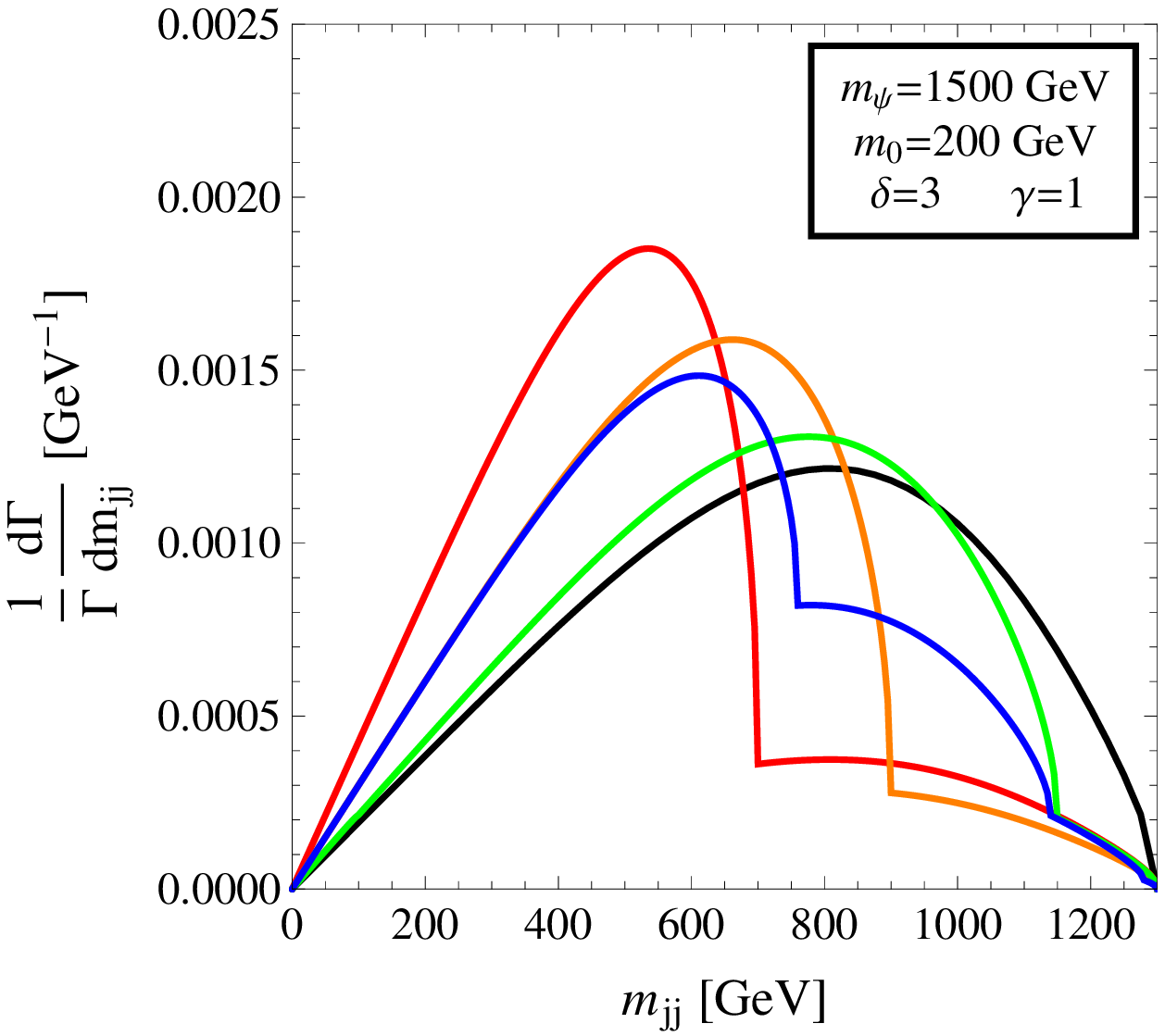}
\end{center}
\caption{DDM invariant-mass distributions, shown for increasing 
$\delta$ with fixed $m_\psi = 1500$~GeV, $m_0=200$~GeV, and $\gamma=1$.  The results
shown in the left, center, and right panels correspond to
$\delta=\{0.5,1,3\}$, respectively.  In each panel, the red, orange, green, and 
blue curves correspond to mass splittings $\Delta m = \{600,400,150,20\}$~GeV, 
respectively, while 
the black curve shows the result for a traditional dark-matter candidate with 
$m_\chi = m_0$. 
\label{fig:mijDistsScaleDelta}}
\begin{center}
  \epsfxsize 2.25 truein \epsfbox {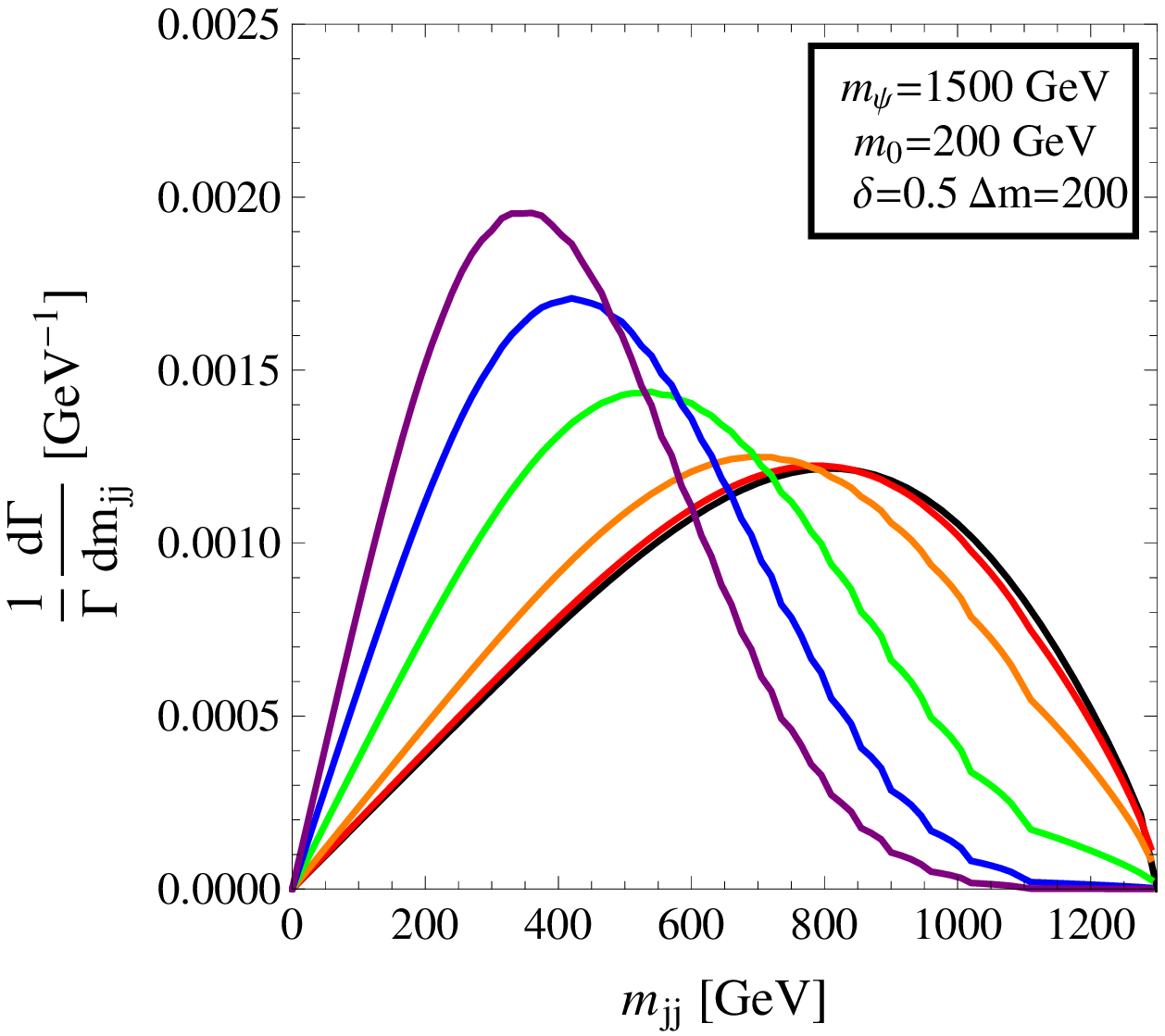}~~
  \epsfxsize 2.25 truein \epsfbox {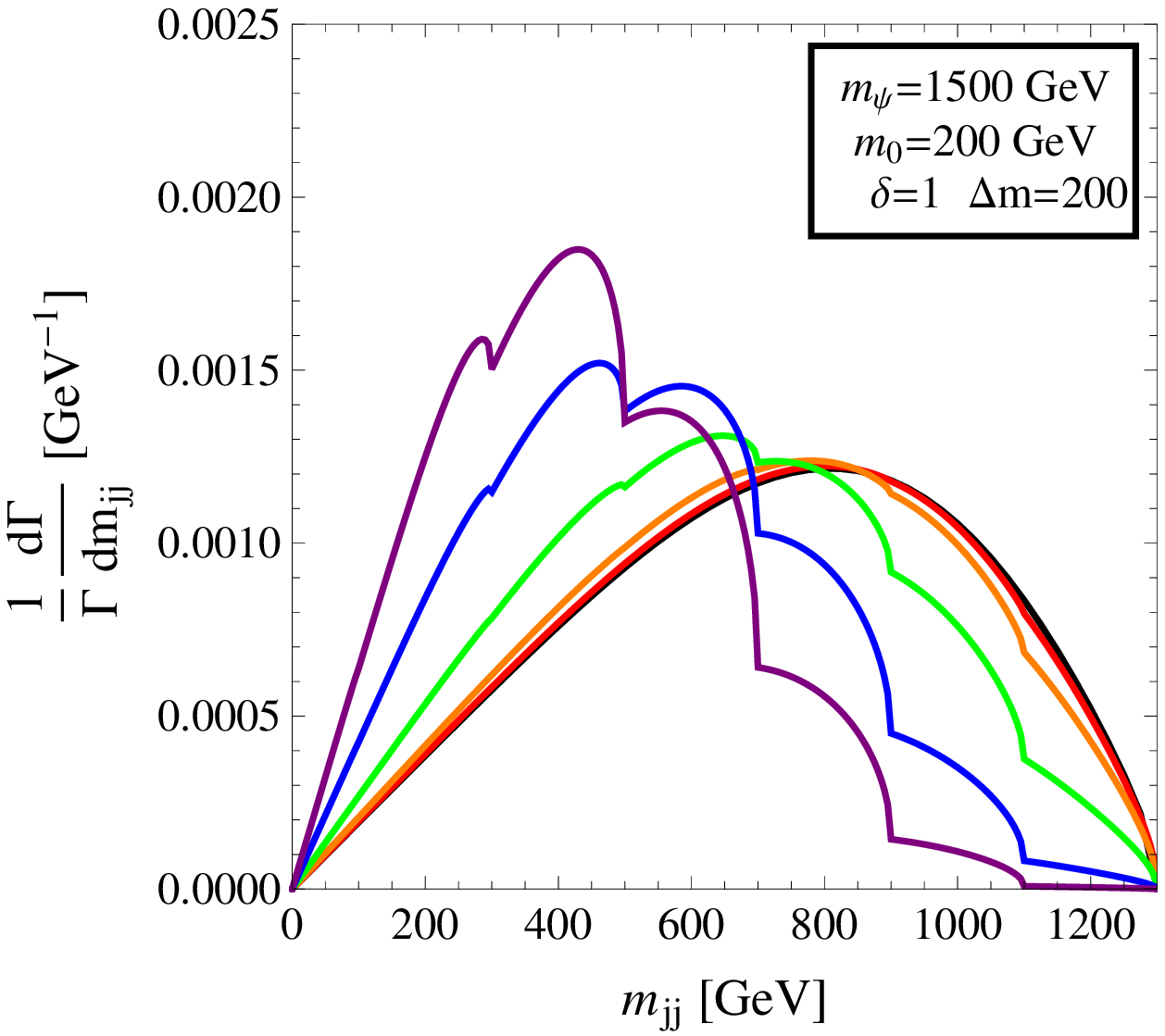}~~
  \epsfxsize 2.25 truein \epsfbox {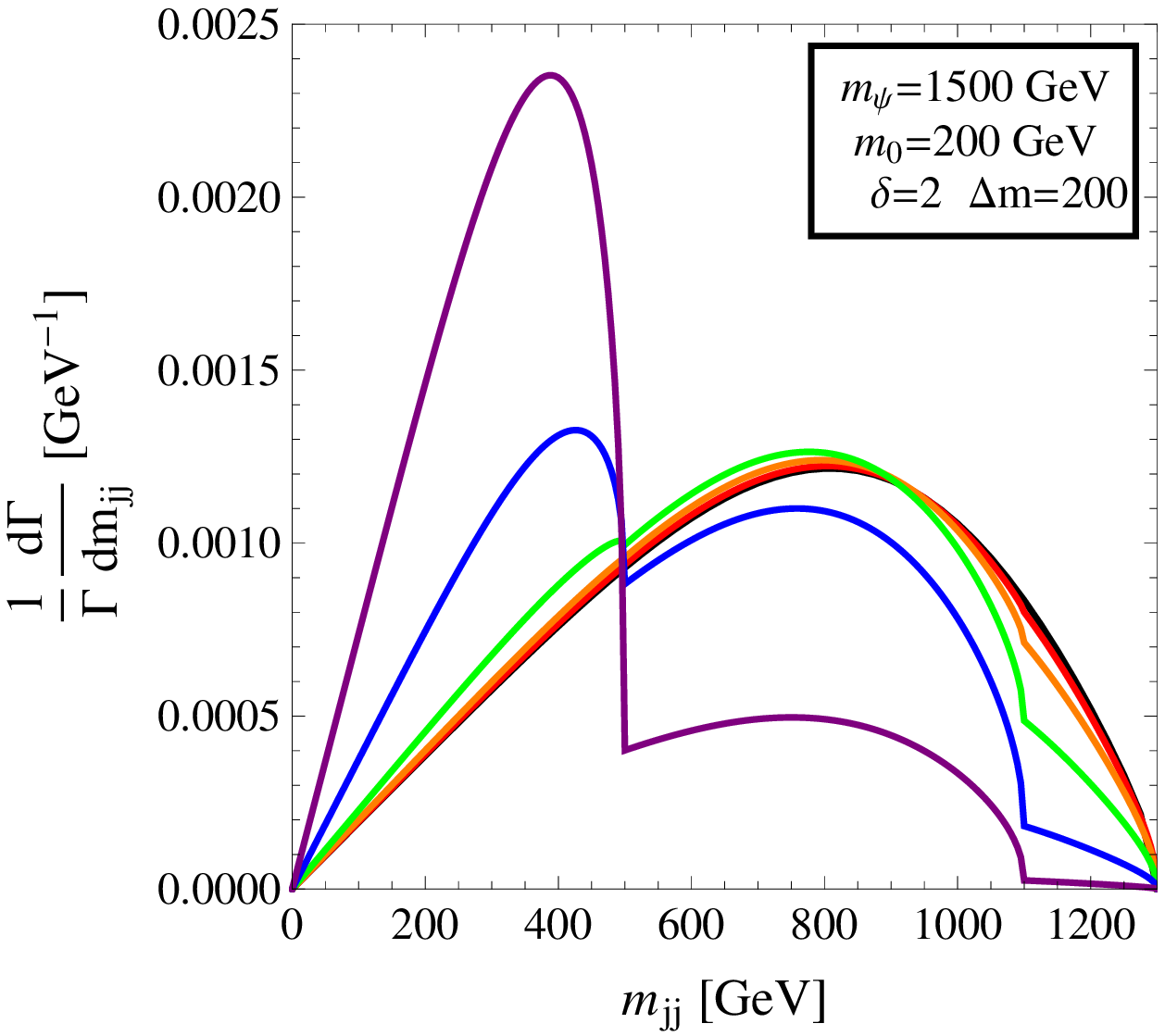}
\end{center}
\caption{DDM invariant-mass distributions, shown for increasing 
$\delta$ with fixed $m_\psi = 1500$~GeV, $m_0=200$~GeV, and $\Delta m = 200$~GeV.  
From top left to bottom right, the panels shown show results for 
$\delta=\{0.5,1,2\}$.  In each panel, the red, orange, green, blue, and 
purple curves correspond to $\gamma=\{-2,-1,0,1,2\}$, respectively, while
the black curve shows the result for a traditional dark-matter candidate with 
$m_\chi = m_0$.  
\label{fig:mijDistsScaleDeltaCurvesDiffGamma}}
\end{figure}

Having examined how the branching fractions $\mathrm{BR}_{\psi n}$ depend on the
parameters $\Delta m$, $\delta$, and $\gamma$ which characterize our DDM ensemble,  
we now turn to examine how the $\mij$ distributions themselves depend on these
parameters.  In Fig.~\ref{fig:mijDistsScaleGamma}, we illustrate how
these distributions depend on $\gamma$ and $\Delta m$ for fixed  $m_\psi = 1500$~GeV, 
$m_0 = 200$~GeV, and $\delta = 1$.  The results shown in the left, center, and right
panels correspond to $\gamma = \{-1,0,1\}$, respectively.     
The red, orange, green, and blue curves in each panel 
respectively correspond to the mass splittings $\Delta m = \{600,400,150,20\}$~GeV, while
the black curve corresponds to the limiting case in which $\Delta m \rightarrow \infty$, or
equivalently to the case of a single dark-matter candidate with mass $m_\chi = m_0$.

It is evident from Fig.~\ref{fig:mijDistsScaleGamma} that in cases in which $\gamma < 0$, 
the heavier fields in the DDM ensemble couple more weakly to $\psi$ than the lighter 
states, to the result that $\chi_0$ dominates in $\Gamma_\psi$.  However, as $\gamma$ 
increases, the $\mathrm{BR}_{\psi n}$ for $n>0$ also increase, to the extent that 
for large $\gamma$, decays to the heavier states in the ensemble actually dominate 
the width of $\psi$.  In this 
latter regime, multiple kinematic edges are evident when $\Delta m$ is sizable.  
While such edges cannot be resolved when $\Delta m$ is small, the peak of the 
distribution nevertheless shifts to smaller values of $\mij$.  This behavior is 
due to the increased branching fractions of $\psi$ to the plethora of heavier 
$\chi_n$ in the DDM ensemble which are kinematically accessible in $\psi$ decays.

In Fig.~\ref{fig:mijDistsScaleDelta}, we illustrate the dependence of the 
$\mij$ distributions on $\delta$ and $\Delta m$ for 
fixed $m_\psi =1500$~GeV, $m_0 = 200$~GeV, and 
$\gamma = 1$.  The left, center, and right panels respectively correspond 
to the cases in $\delta = \{0.5,1,3\}$.  We see from the right panel that for large 
$\delta$, and especially when $\Delta m$ is large, the $\mij$ distribution is sensitive 
primarily to only the lightest few $\chi_n$ in the DDM ensemble.  As a result,
the distributions shown are characterized by the presence of several identifiable 
mass edges, each corresponding to one of these light $\chi_n$.  By contrast, we 
see in the left panel that for 
small $\delta$, a qualitatively different situation emerges:  
in this regime, the individual mass edges become difficult to distinguish, 
especially for small $\Delta m$, and the peaks of the invariant-mass distributions 
shift to lower values of $\mij$.         

Finally, in Fig.~\ref{fig:mijDistsScaleDeltaCurvesDiffGamma}, 
we illustrate the dependence of the 
$\mij$ distributions on $\delta$ and $\gamma$ for fixed $m_\psi =1500$~GeV, 
$m_0 = 200$~GeV, and $\Delta m = 200$~GeV.  The left, center, and right panels 
correspond respective to the cases in which $\delta = \{0.5,1,2\}$.  As in 
Fig.~\ref{fig:mijDistsScaleDelta}, for large $\delta$ we see that the $\mij$ 
distribution is sensitive primarily to the lightest states in the ensemble and 
features an identifiable mass edge corresponding to each such state, while for 
small $\delta$ the mass edges become increasingly indistinct.  
However, we also see from the left panel that the overall shape of the 
distribution and the location of its peak still depend on $\gamma$, even in 
situations in which no individual mass edge can be identified.  Thus, even 
in this limit, we find that the shape of the $\mij$ distribution conveys 
non-trivial information about the structure of the DDM ensemble.

To summarize, we see that DDM models give rise to two 
characteristic classes of $\mij$ distributions in different regimes of 
model parameter space.  Each of these represents a dramatic departure from
the $\mij$ distributions typically realized in traditional dark-matter models. 
In the regime in which only a few states contribute significantly to the 
decay width of the parent particle $\psi$, the $\mij$ distribution is
characterized by the presence of multiple identifiable mass edges. 
This occurs either when the mass splittings between the heavier states in the 
ensemble are large, due to either $\Delta m$ or $\delta$ being sizeable.  
To some extent, this behavior is not unexpected, but the presence of such 
identifiable mass edges may ultimately provide one clear way of discerning a DDM 
ensemble experimentally.  By contrast, in the opposite regime in which a 
large number of states contribute significantly to the decay width of the 
parent particle $\psi$, the $\mij$ distribution exhibits no identifiable
edges, but instead assumes a unique shape which is markedly different from that 
observed in traditional dark-matter models.  Such a shape emerges when mass 
splittings between the heavier states in the ensemble are small, and consequently 
a larger fraction of that ensemble (consisting primarily of the heavier $\chi_n$) 
contributes non-trivially to the decay width of $\psi$.


\section{Distinguishing DDM Ensembles at the LHC\label{sec:significances}}


In the previous section, we examined a number of distinctive features which can 
emerge in the invariant-mass distributions associated with $\psi\rightarrow j j\chi_n$ 
decays in DDM models.  In this section, our primary aim is to assess 
the degree to which the characteristic $\mij$ distribution associated with a 
particular DDM model constitutes a distinctive signature of non-standard physics 
in the dark sector --- \ie, a signature that cannot be realized in any 
traditional dark-matter model, regardless of the mass $m_\chi$ of the 
dark-matter candidate.  Once again, for concreteness, 
we focus on the operator structure given in Eq.~(\ref{eq:FourFermiOp}).    
In order to determine the statistical significance with which the 
$\mij$ distribution associated with any particular DDM model 
(\ie, a particular set of values for the parameters 
$m_\psi$, $m_0$, $\Delta m$, $\gamma$, $\delta$) is truly distinctive, 
we compare this distribution to the $\mij$ distributions 
associated with a variety of different traditional dark-matter models, each 
with a different value of $m_\chi$.  In particular, we canvass the entire range
$0 < m_\chi < m_\psi - \mijmin$ with a finite step size.  
Note that we do {\it not} similarly scan over all possible spins combinations for 
$\psi$ and $\chi$, or over coupling structures of these particles to SM quarks and 
gluons, but rather restrict our analysis to traditional dark-matter models in which 
the parent particle likewise decays via an operator of the form given in
Eq.~(\ref{eq:FourFermiOpStdDM}).  This is justified because the shape of the 
$\mij$ distribution does not depend sensitively on the coupling structure, 
as we have demonstrated in Sect.~\ref{sec:mijDistributions}.
We have nevertheless verified that our results do not differ significantly even if 
alternative coupling structures are incorporated into the analysis.      
  
The procedure we adopt in comparing any two $\mij$ distributions is as follows.
When $\mij$ is small, statistics are low and residual SM backgrounds are at their 
largest; we therefore begin by applying a minimum cut on $\mij$ of the form 
$\mij > \mijmin$, where $\mijmin$ is some particular invariant-mass threshold.         
We then partition the range of allowed dijet invariant masses 
$\mijmin \leq \mij \leq m_\psi - m_0$ into $n_b$ bins whose widths vary with
$\mij$.  Specifically, the width of each bin is taken to be equal to the dijet 
invariant-mass resolution $\Delta \mij$ at the minimum $\mij$ in the bin.
We assume here that $\Delta \mij$ is limited predominately by the uncertainty 
$\Delta E_j$ in the measurement of the energies $E_j$ of the jets used in 
reconstructing $\mij$, and hence we take $(\Delta \mij)/\mij \approx (\Delta E_j)/E_j$. 
For the ATLAS detector, the jet-energy resolution 
has the rough scaling behavior~\cite{ATLASTDR}  
\begin{equation}
  \frac{\Delta E_j}{E_j} ~=~ 0.5 \left(\frac{E_j}{\mathrm{GeV}}\right)^{-1/2} + 0.03~,
\end{equation}
and the jet-energy resolution at CMS is quite similar~\cite{CMSDijetSearch}.

For each value of $m_\chi$ in our survey, we assess the goodness of fit between the 
$\mij$ distributions associated with the specific DDM model under study and 
traditional dark-matter models by constructing the $\chi^2$ statistic 
\begin{equation}
  \chi^2(m_{\chi}) ~=~ \sum_k\frac{[X_k - \mathcal{E}_k(m_\chi)]^2}{\sigma_k^2}~,
\end{equation}
where the index $k$ labels the bin, $X_k$ is the expected population of events in 
bin $k$ in the DDM model, $\mathcal{E}_k(m_\chi)$ is the expected population of events in
bin $k$ in a traditional dark-matter model in which the dark-matter particle 
has mass $m_\chi$, and $\sigma_k^2$ is the variance in $X_k$ due to statistical 
uncertainties.  Since the $X_k$ are distributed according to a multinomial 
distribution, it follows that $\sigma_k^2 = X_k (1-X_k/N_e)$, where $N_e$ denotes 
the total number of signal events in the sample.  Note that the 
minimum from among these $\chi^2(m_\chi)$ values embodies the degree of discrepancy 
between the $\mij$ distribution associated with the DDM model and that associated 
with the traditional dark-matter model which provides the best fit to that distribution. 
We therefore adopt 
\begin{equation}
  \chi^2_{\mathrm{min}} ~\equiv~ \min_{m_\chi} \big\{\chi^2(m_\chi)\big\}
  \label{eq:ChiSqMinDef}
\end{equation}
as our final measure of the distinctiveness of the $\mij$ distribution associated with 
the DDM model.  We further determine the statistical significance associated 
with a particular value of $\chi^2_{\mathrm{min}}$ by first comparing this value to 
a $\chi^2$ distribution with $n_b - 1$ degrees of freedom in order to obtain 
a $p$-value.  We then take the significance level to be the number of standard 
deviations away from the mean to which the same $p$-value would correspond for a 
Gaussian distribution.  Note that unlike in typical bump-hunting analyses, 
the $p$-value in this case is two-sided.
 
\begin{figure}[t!]
\begin{center}
  \epsfxsize 2.25 truein \epsfbox
    {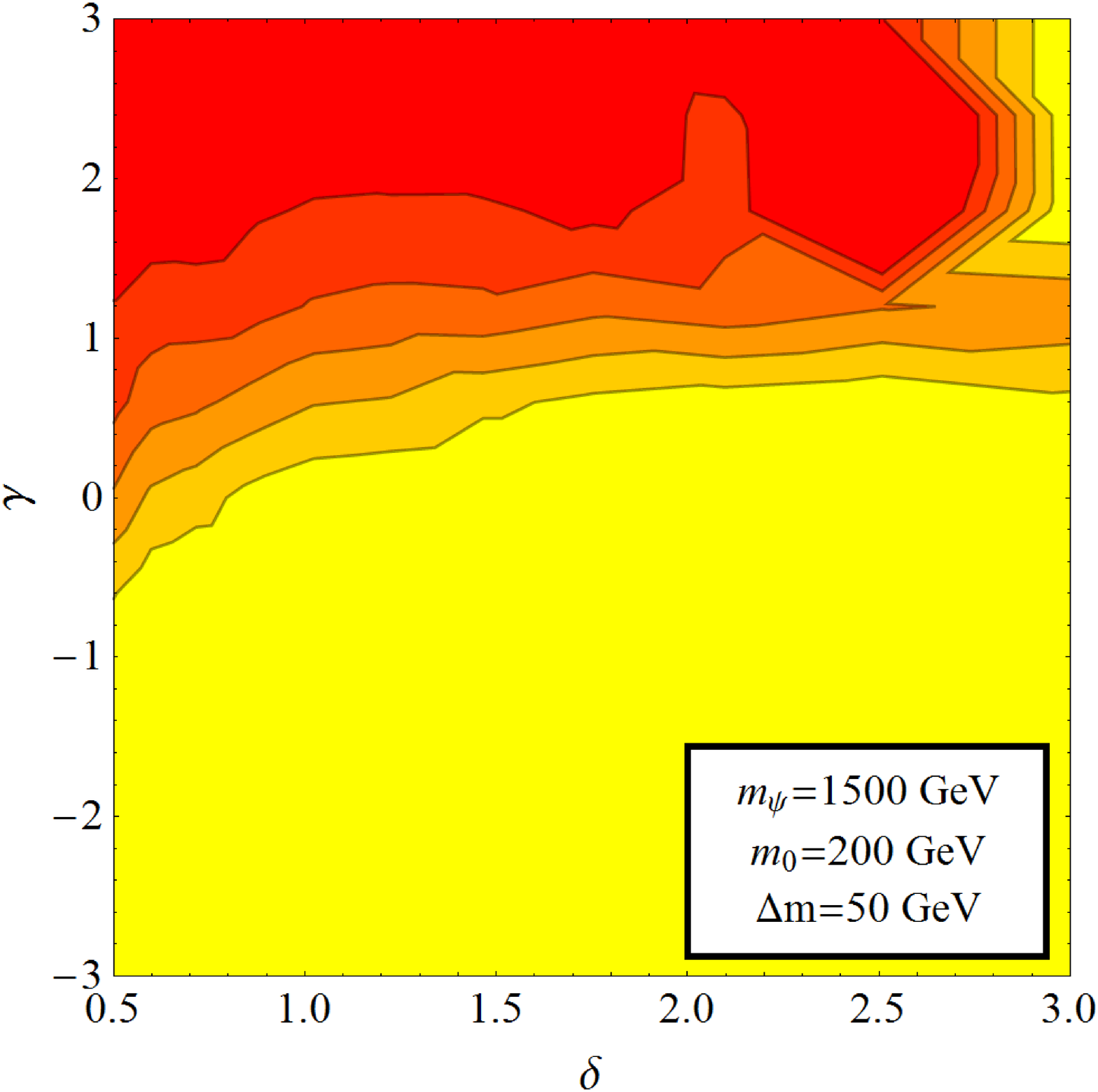}
  \epsfxsize 2.25 truein \epsfbox
    {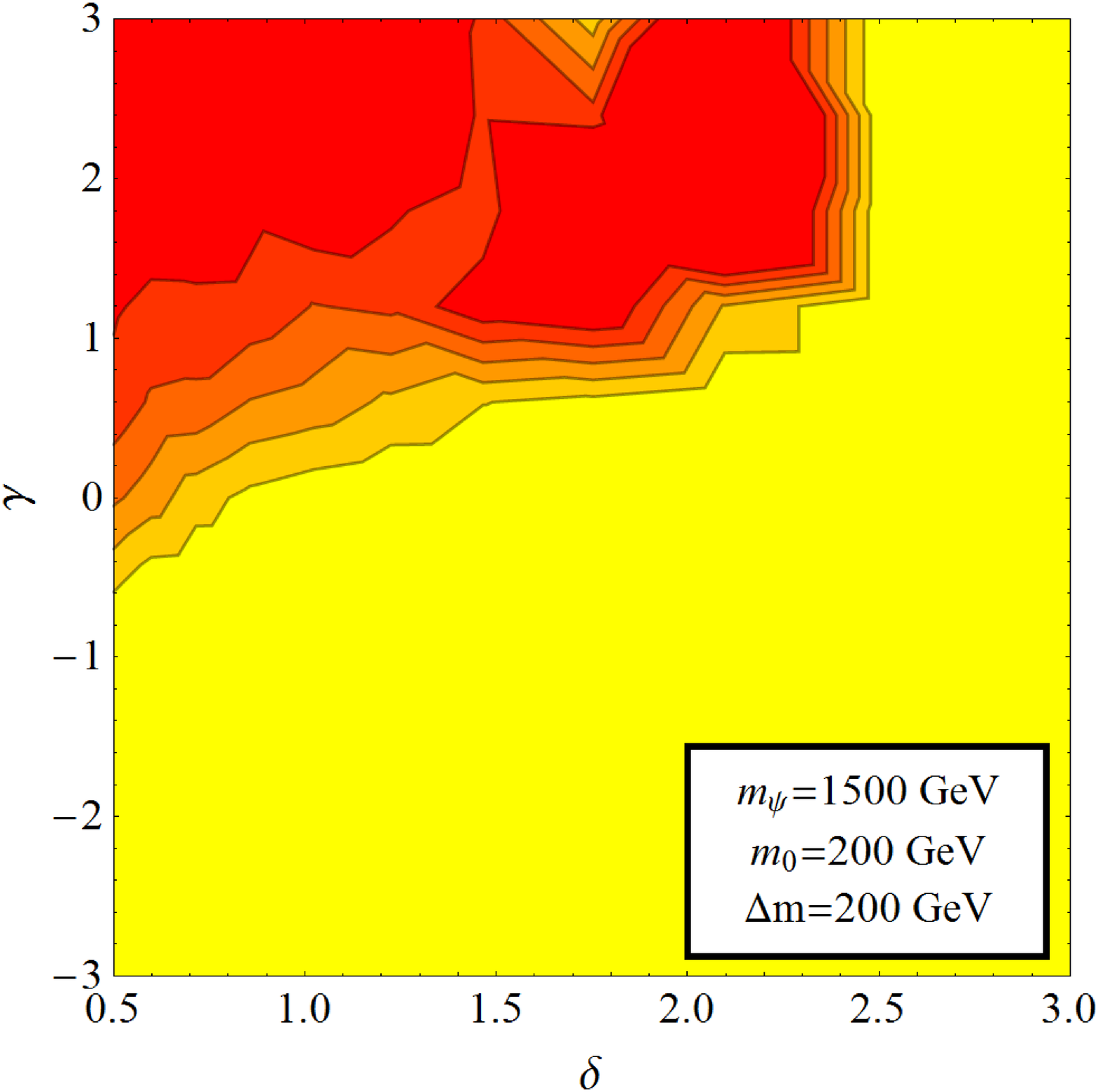}
  \epsfxsize 2.25 truein \epsfbox
    {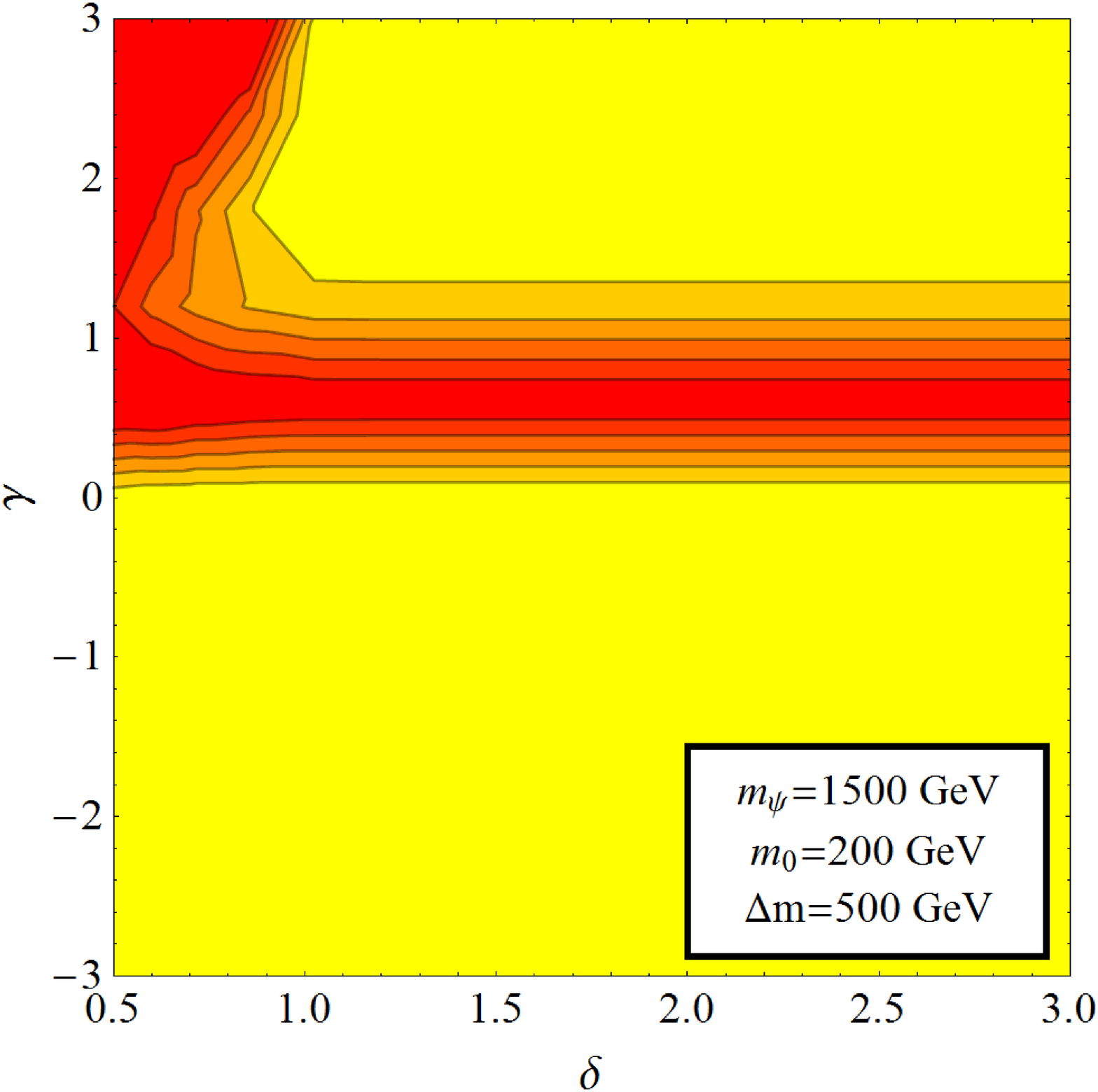}\\
  \epsfxsize 2.25 truein \epsfbox 
    {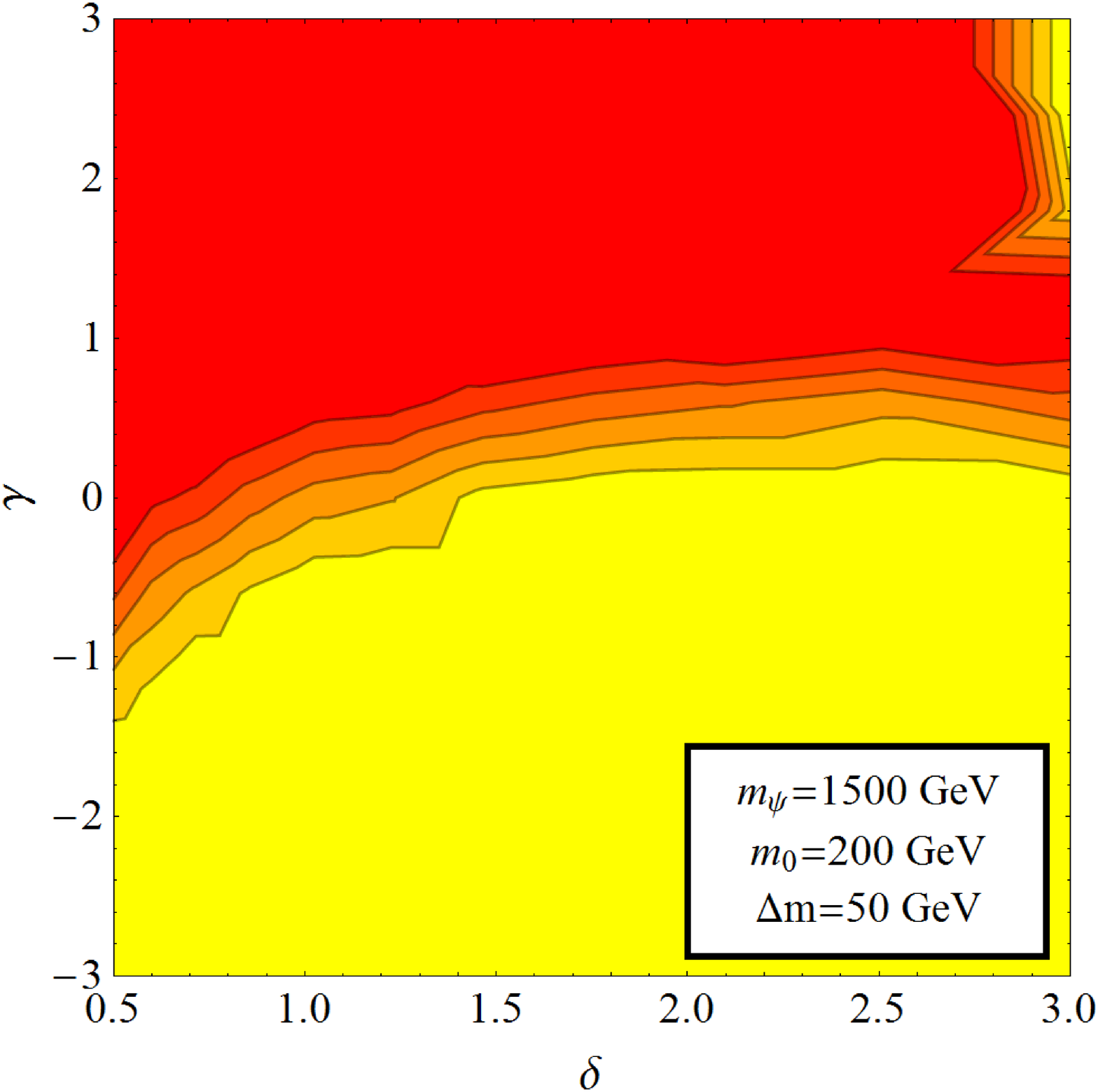}
  \epsfxsize 2.25 truein \epsfbox
    {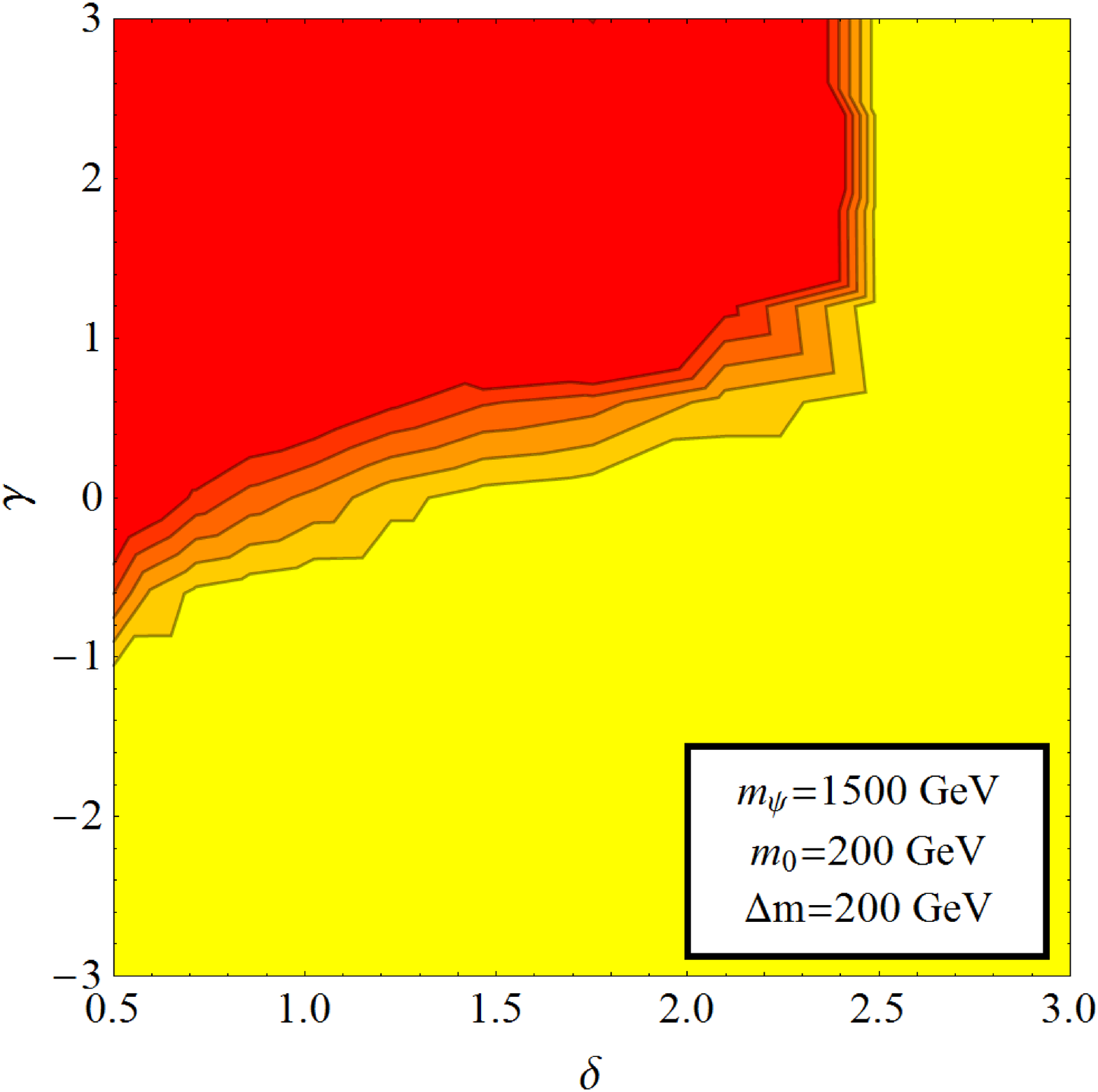}
  \epsfxsize 2.25 truein \epsfbox
    {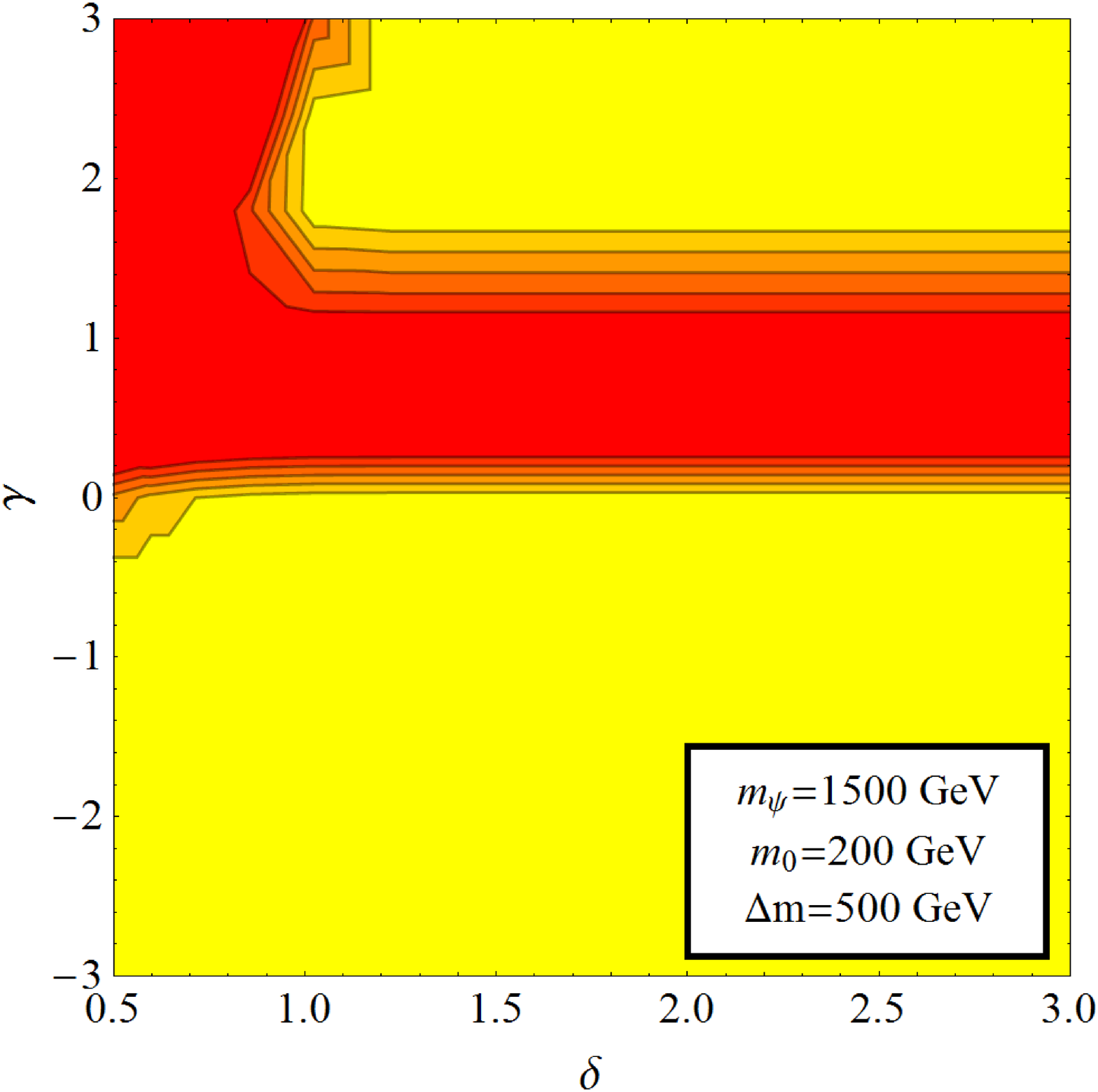}
  \raisebox{0.5cm}{\large Significance:~~}
     \epsfxsize 5.00 truein \epsfbox {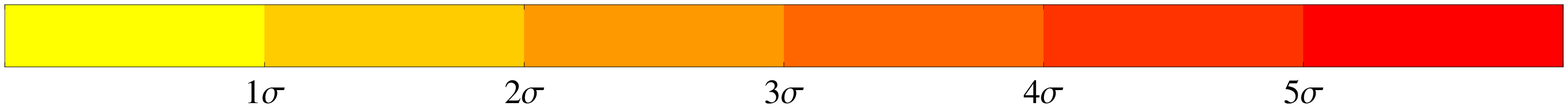}
\end{center}
\caption{Contour plots showing the minimum significance 
level at which the $\mij$ distribution predicted in the DDM model characterized 
by the parameters $m_\psi$, $m_0$, $\Delta m$, $\delta$, and $\gamma$ 
can be differentiated from the $\mij$ distribution predicted in any traditional 
dark-matter model.  In the panels shown, we have fixed $m_\psi = 1500$~GeV 
and $m_0 = 200$~GeV, and taken $\mijmin = 200$~GeV as our 
minimum $\mij$ threshold.  The panels in the left column show results 
for $\Delta m = 50$~GeV, the center column for $\Delta m = 200$~GeV,
and the right column for $\Delta m = 500$~GeV.  In each column, the top 
panel corresponds to $N_e = 500$, while the bottom panel
corresponds to $N_e = 1000$.
\label{fig:SurveysFixedDeltam}}
\end{figure}
  
In Fig.~\ref{fig:SurveysFixedDeltam}, we show how the significance of differentiation 
varies as a function of $\delta$ and $\gamma$ with $\Delta m$, $m_0$, and $m_\psi$ 
held fixed.  In all of the panels shown, we have set $m_\psi = 1500$~GeV and 
$m_0 = 200$~GeV, and we have taken $\mijmin = 200$~GeV as 
our minimum $\mij$ threshold and a step size of $100$~GeV in our scan over $m_\chi$.
The panels in the left column 
show the results for $\Delta m = 50$~GeV, the middle column for $\Delta m = 200$~GeV,
and the right column for $\Delta m = 500$~GeV.  In each column, the results shown in the 
top and bottom panels correspond to $N_e = 500$ and $N_e = 1000$, 
respectively.  

The results in the left column indicate that
for relatively small $\Delta m$, our DDM model can be distinguished from traditional 
dark-matter scenarios at a significance level of $5\sigma$ or higher for 
$\gamma \gtrsim 0.5$ and $\delta \lesssim 3$ with $N_e \gtrsim 1000$.     
This is due to the fact that the coupling to the heavier $\chi_n$ are proportionally 
larger for $\gamma > 0$, and these states increasingly dominate the width of $\psi$ as
$\gamma$ increases.  As a result, the peak of the $\mij$ distribution shifts to smaller 
smaller values of $\mij$, and is therefore more readily distinguished from the 
distribution obtained in traditional dark-matter models.  By contrast, for $\gamma < 0$,  
the width of $\psi$ is dominated by decays to the lighter $\chi_n$, and the $\mij$ 
distributions obtained in this regime resemble much more closely those associated with
traditional dark-matter models.  This behavior is evident in the center and right panels
as well.  

We also see from the panels of Fig.~\ref{fig:SurveysFixedDeltam} that 
the significance for differentiation depends non-trivially on $\delta$ and 
$\Delta m$ as well.
For example, an increase in significance within the $\gamma < 0$ region is obtained 
for small $\delta$ --- an increase which is particularly pronounced for small 
$\Delta m$.  This effect is a consequence of the sheer multiplicity of heavy $\chi_n$ 
overwhelming the effect of the coupling suppression of $\psi$ to these states (which 
occurs for $\gamma < 0$) and resulting in a $\mij$ distribution which peaks at far 
smaller values of $\mij$ than in traditional dark-matter models.  Conversely, in the 
opposite regime in which $\gamma > 0$ and $\delta$ is large, a precipitous 
drop in the significance occurs for $\delta \gtrsim 3$ in the $\gamma \gtrsim 0.5$ 
region for $\Delta m = 50$~GeV, and at even smaller values of $\delta$ for larger 
$\Delta m$.  This behavior reflects the fact that as $\Delta m$ and $\delta$ 
increase, the mass splittings between the $\chi_n$ eventually become larger, and 
consequently fewer and fewer $\chi_n$ remain kinematically accessible in $\psi$ decays.         
In such cases, the contribution from either of these states can mimic the $\mij$ 
distribution obtained for a single dark-matter particle, depending on 
the branching fractions $\mathrm{BR}_{\psi 0}$ and $\mathrm{BR}_{\psi 1}$.  Whenever 
$\mathrm{BR}_{\psi 0} \gg \mathrm{BR}_{\psi 1}$ or
$\mathrm{BR}_{\psi 0} \ll \mathrm{BR}_{\psi 1}$, only one of these two 
particles dominates the width of $\psi$, and the resulting $\mij$ distribution 
resembles that obtained for a traditional dark-matter candidate with $m_\chi = m_0$ or
$m_\chi = m_1$, respectively.  However, in situations in which
$\mathrm{BR}_{\psi 0} \sim \mathrm{BR}_{\psi 1}$ and
both particles have a salient effect on the $\mij$ distribution, it takes a shape which 
no single-particle dark-matter candidate can replicate.  Since the relationship between 
these branching fractions is primarily governed by $\gamma$, 
there exists a narrow range $0\lesssim \gamma \lesssim 1$ within which 
the significance for differentiation is substantial for large $\delta$ and 
$\Delta m$.  By contrast, outside of this region, either $\chi_0$ or $\chi_1$ 
overwhelmingly dominates the width of $\psi$, and the significance drops 
dramatically.  Indeed, this behavior is evident in the right 
panels of Fig.~\ref{fig:SurveysFixedDeltam}.

The plots in each column of Fig.~\ref{fig:SurveysFixedDeltam} illustrate 
the effect of $N_e$ on the significance of differentiation, and in particular on
the sensitivity of our results to the total event count.  Indeed, we 
observe that a $5\sigma$ significance of differentiation is achieved over a 
a substantially larger region of parameter space for $N_e = 1000$ than for 
$N_e = 500$.  Moreover, it turns out that the region of parameter space 
in which such a significance is obtained diminishes rapidly with decreasing 
$N_e$ for $N_e \lesssim 500$.  Thus, we conclude that
$N_e \sim \mathcal{O}(500 - 1000)$ is roughly the size of the data sample 
for which $\mij$ distributions begin to provide significant resolving power
between DDM scenarios and traditional dark-matter models.  
In Sect.~\ref{sec:productionchannels}, we shall provide an example of a 
production mechanism which naturally yields such $N_e$ values for 
$\Lint = 30\mathrm{~fb}^{-1}$ at the $\sqrt{s} = 14$~TeV LHC.

\begin{figure}[ht!]
\begin{center}
  \epsfxsize 2.25 truein \epsfbox
    {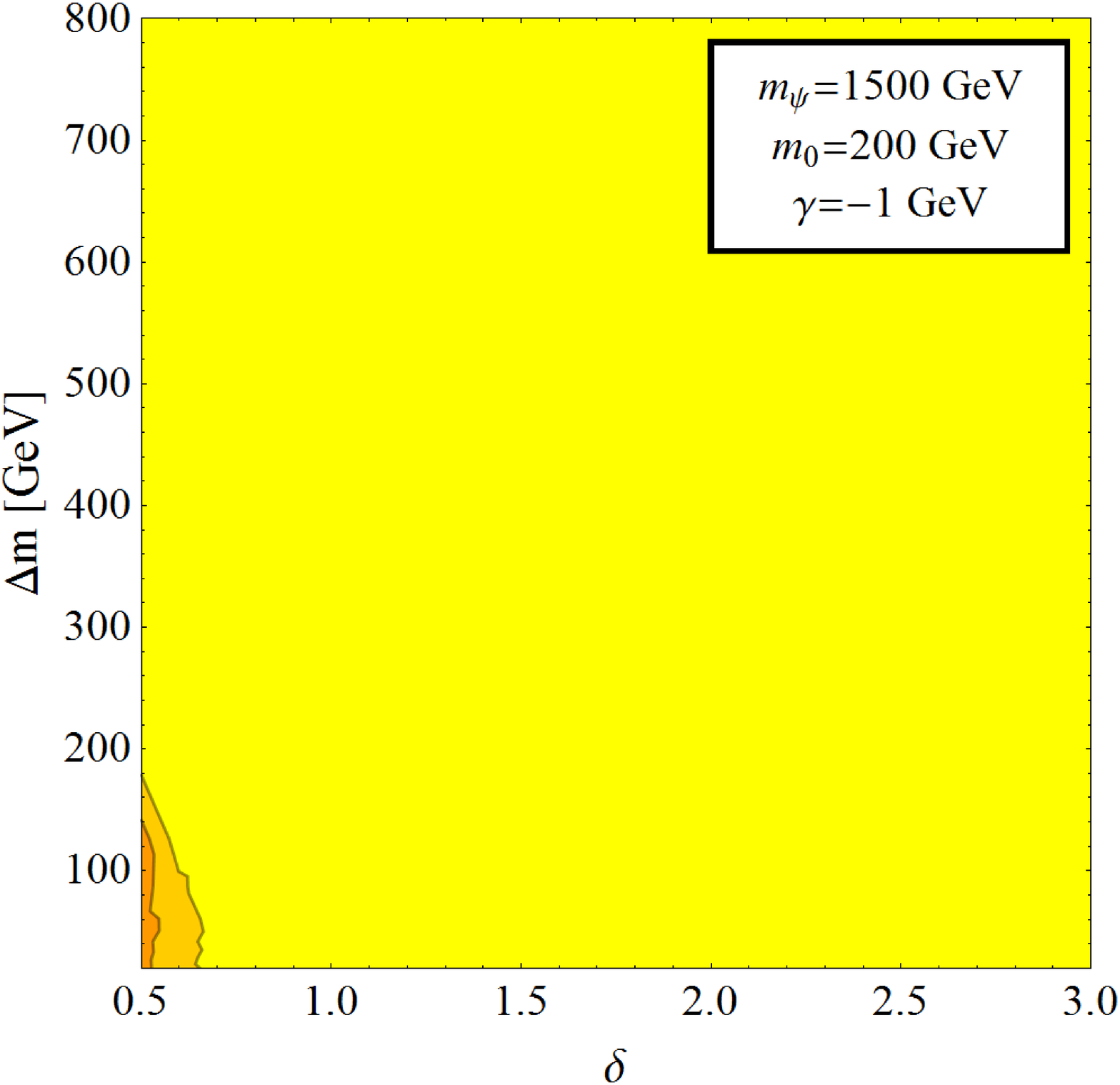}
  \epsfxsize 2.25 truein \epsfbox 
    {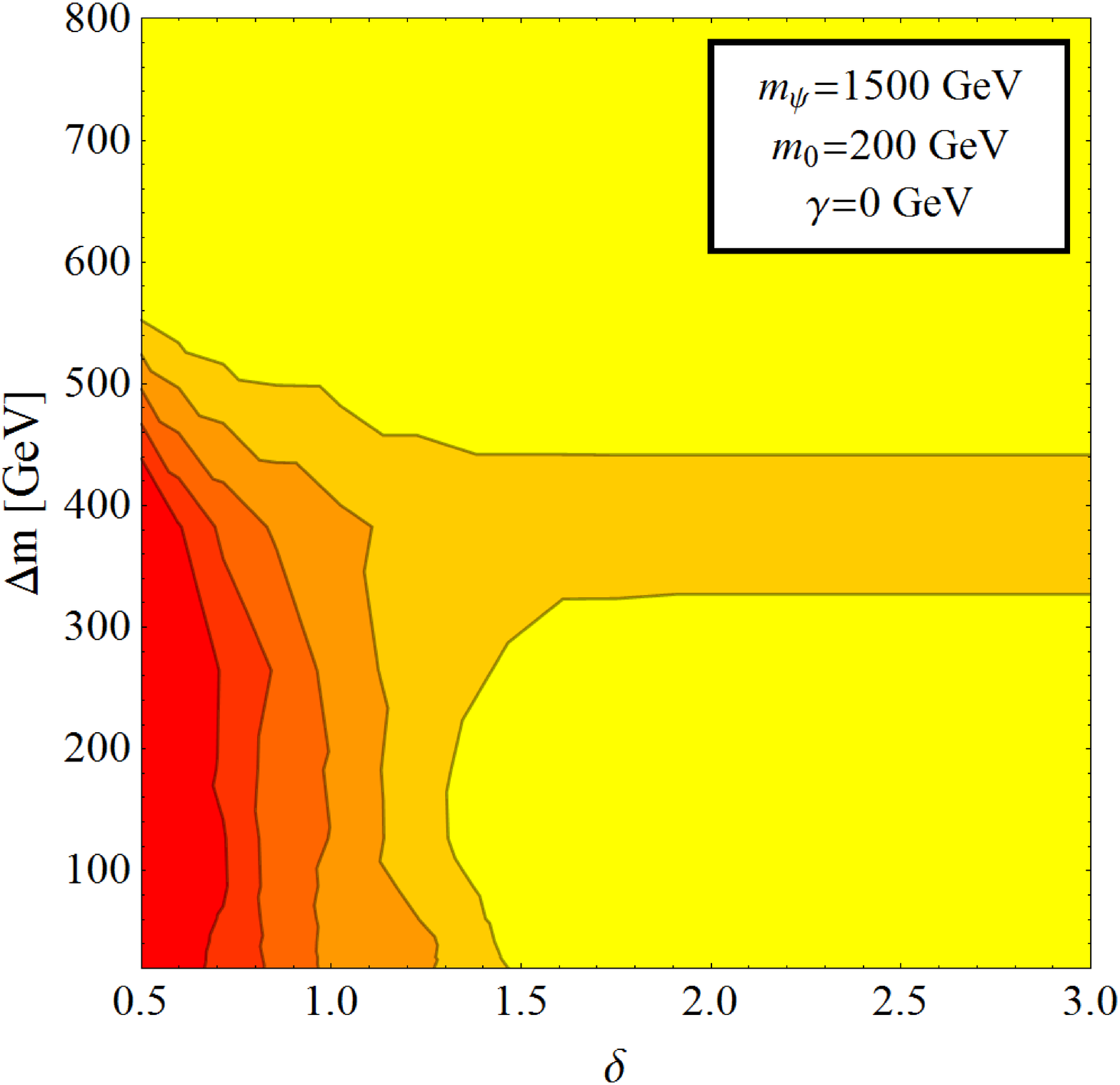}
  \epsfxsize 2.25 truein \epsfbox
    {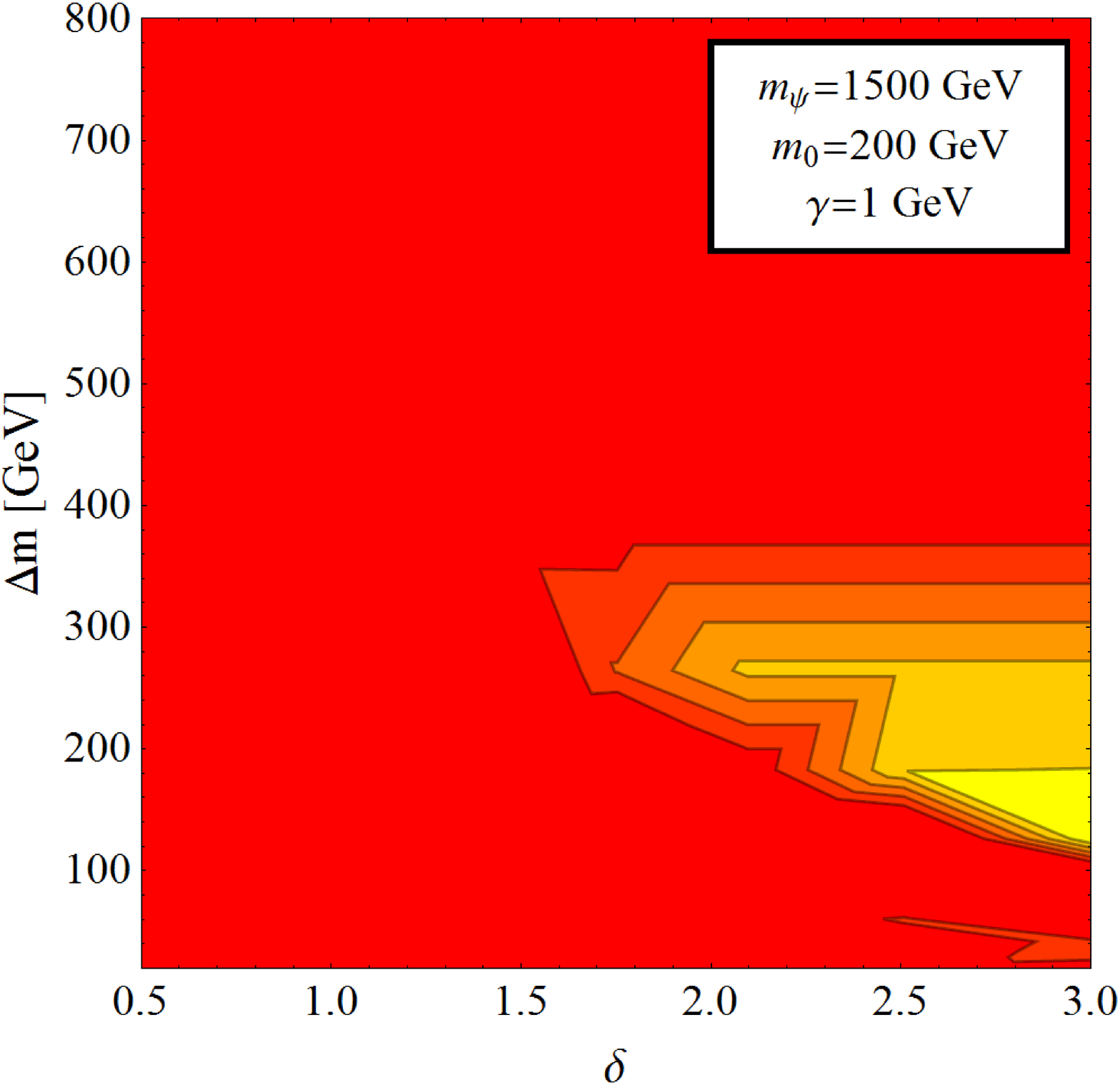}
\end{center}
\caption{
Contour plots showing the minimum significance 
level at which the $\mij$ distribution predicted in the DDM model characterized 
by the parameters $m_\psi$, $m_0$, $\Delta m$, $\delta$, and $\gamma$ 
can be differentiated from the $\mij$ distribution predicted in any traditional 
dark-matter model for $N_e = 1000$. 
The colored regions shown correspond to the same significance intervals as in 
Fig.~\protect\ref{fig:SurveysFixedDeltam}, and the results shown correspond to the
same values of $m_\psi$, $m_0$, and $\mijmin$.  The left panel shows results 
for $\gamma = -1$, the center panel for $\gamma = 0$,
and the right panel for $\gamma = 1$.
\label{fig:SurveysFixedGamma}}
\vskip 0.325 truein
\begin{center}
  \epsfxsize 2.25 truein \epsfbox
    {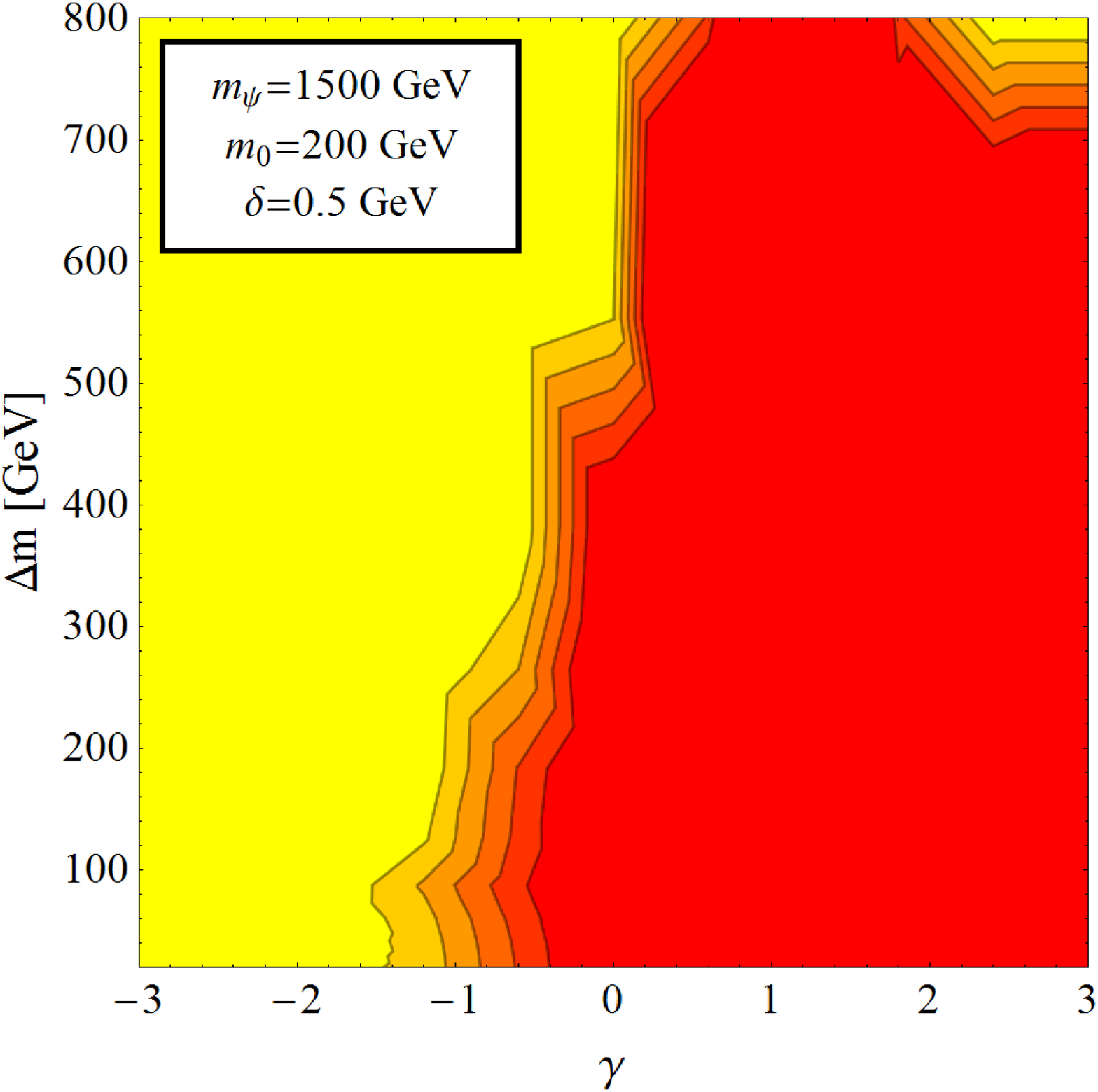}
  \epsfxsize 2.25 truein \epsfbox 
    {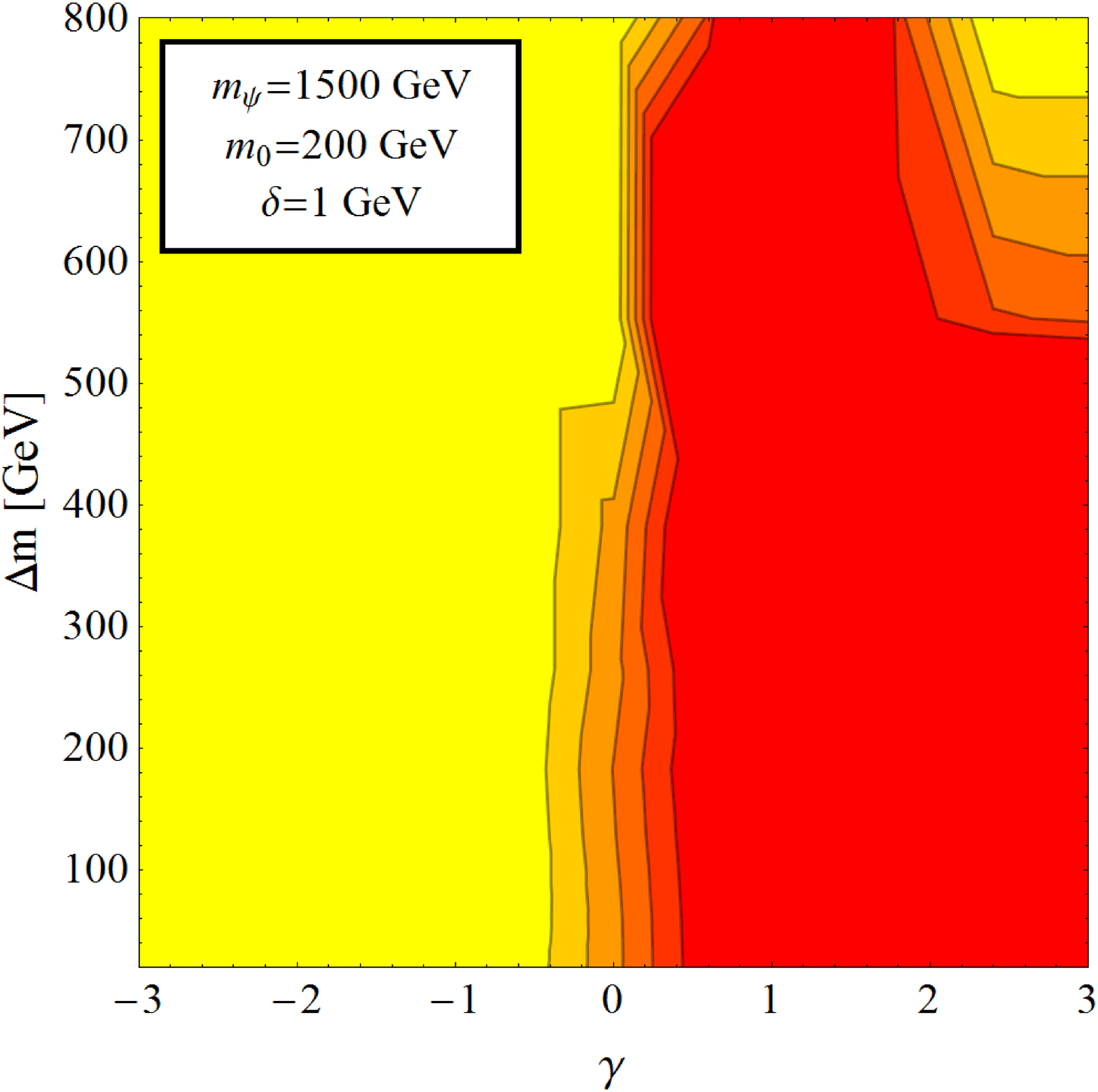}
  \epsfxsize 2.25 truein \epsfbox
    {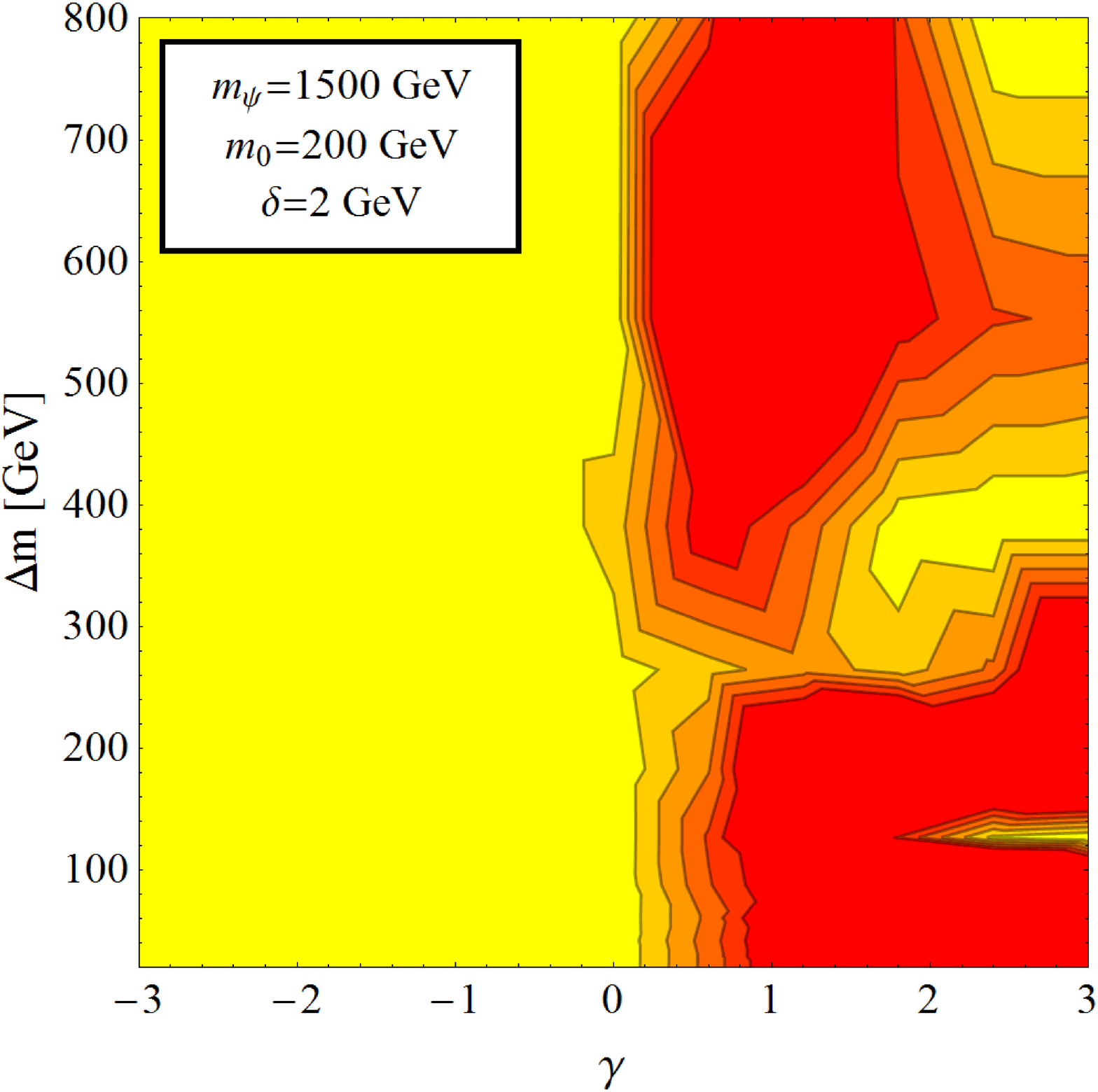}
\end{center}
\caption{Contour plots showing the minimum significance 
level at which the $\mij$ distribution predicted in the DDM model characterized 
by the parameters $m_\psi$, $m_0$, $\Delta m$, $\delta$, and $\gamma$ 
can be differentiated from the $\mij$ distribution predicted in any traditional 
dark-matter model for $N_e = 1000$.
The colored regions shown correspond to the same significance intervals as in 
Fig.~\protect\ref{fig:SurveysFixedDeltam}, and the results shown correspond to the
same values of $m_\psi$, $m_0$, and $\mijmin$.  The left panel shows results 
for $\delta = 0.5$, the center panel for $\delta = 1$,
and the right panel for $\delta = 2$.
\label{fig:SurveysFixedDelta}}
\end{figure}  

In Fig.~\ref{fig:SurveysFixedGamma}, we show how the significance of differentiation 
varies as a function of $\Delta m$ and $\delta$ with $\gamma$, $m_0$, and $m_\psi$ 
held fixed.  The results shown correspond to the same choices of   
$m_\psi$, $m_0$, $\mijmin$, and step size for $m_\chi$ as in
Fig.~\ref{fig:SurveysFixedDeltam}.  The left, center, and right panels show the results 
for $\gamma = \{-1,0,1\}$, respectively.  Once again,
we see from the left panel that for $\gamma < 0$, the density of states 
as a function of mass within the DDM ensemble must increase quite rapidly with 
$n$ to overcome the corresponding coupling suppression.  Both 
$\Delta m$ and $\delta$ must be extremely small for this to occur.  As $\gamma$ 
increases, less extreme values for these parameters are required to overcome 
the coupling suppression, as the results shown in the center panel of the figure 
attest.  A horizontal band of slightly increased significance for   
$300\mathrm{~GeV}\lesssim\Delta m\lesssim 450\mathrm{~GeV}$ is also evident
for $\delta \gtrsim 1.5$.  This once again corresponds to the region of parameter 
space discussed above, within which $\chi_0$ and $\chi_1$ are the only states in 
the ensemble kinematically accessible in $\psi$ decays, but within which the
$\mij$ distribution nevertheless differs significantly from those associated with
traditional dark-matter candidates because  
$\mathrm{BR}_{\psi 0} \sim \mathrm{BR}_{\psi 1}$.
In the right panel, the couplings of the heavier $\chi_n$ are enhanced 
relative to those of their lighter counterparts, and as a result, the only 
regions of parameter space shown in which a $5\sigma$ significance is not obtained
are those which correspond to the upper right portions of the $\Delta m =50$~GeV and
$\Delta m =200$~GeV panels in Fig.~\ref{fig:SurveysFixedDeltam}, within which only
two of the $\chi_n$ are kinematically accessible in $\psi$ decays, and one of these 
two particles overwhelmingly dominates the width of $\psi$. 

Finally, in Fig.~\ref{fig:SurveysFixedDelta}, we show how the significance of
differentiation varies as a function of $\Delta m$ and $\gamma$ with $\delta$, 
$m_0$, and $m_\psi$ held fixed.  Once again,
the results shown correspond to the same choices of   
$m_\psi$, $m_0$, $\mijmin$, and step size for $m_\chi$ as in
Fig.~\ref{fig:SurveysFixedDeltam}.  The left, center, and right panels show the results 
for $\delta = \{0.5,1,2\}$, respectively.  The results 
in these panels illustrate once again that large $\gamma$ generally yields a large
significance for differentiation --- except in situations in which $\chi_0$ and 
$\chi_1$ are the only $\chi_n$ sufficiently light to be produced by $\psi$ decays, and
the condition $\mathrm{BR}_{\psi 0} \sim \mathrm{BR}_{\psi 1}$ 
must be satisfied in order for the $\mij$ distribution to differ significantly from 
that of a traditional dark-matter model with either $m_\chi = m_0$ or $m_\chi = m_1$. 
   
Taken together, the results displayed in Figs.~\ref{fig:SurveysFixedDeltam}
through~\ref{fig:SurveysFixedDelta} indicate that it is possible to discriminate  
between a wide range of DDM ensembles and more traditional dark-matter candidate 
on the basis of dijet invariant-mass distributions at the LHC.  Indeed, we have shown 
that a statistical significance for discrimination in excess of $5\sigma$ is obtained for
$N_e = 1000$ over a broad region of the DDM model parameter space, and
regions of parameter space within which $\gamma$ is large or within 
which $\Delta m$ is small and $\delta < 1$ are particularly auspicious.


\section{DDM Production Channels and Event Rates\label{sec:productionchannels}}

  
Thus far, we have not specified any particular production mechanism for the 
parent particle $\psi$ whose decays give rise to the constituent fields 
of the DDM ensemble.  Indeed, all of our results up to this point  
have been essentially independent of the details of the mechanism by 
which the $\psi$ are produced, provided that the SM 
background can be reduced to a negligible level by cuts imposed on the data 
and that jets associated with the decay of each parent particle can be correctly 
identified.  Indeed, they depend only on the total event count $N_e$.  
This gives our analysis a broad range of applicability.  We have 
demonstrated that in a wide variety of situations, 
$N_e \sim \mathcal{O}(500 - 1000)$ is sufficient to 
distinguish a DDM ensemble from a traditional dark-matter candidate at the 
$5\sigma$ significance level.  
It is therefore natural to wonder whether  
such values of $N_e$ might reasonably be expected within the next few years of 
operation at the LHC, and what sorts of parent-particle production processes 
are capable of yielding such event counts.

\begin{figure}[ht!]
\centerline{
  \epsfxsize 2.0 truein \epsfbox {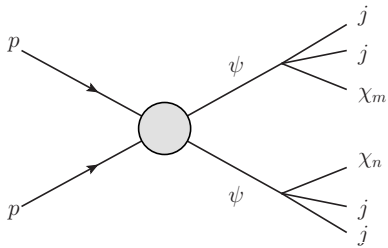}  }
\caption{One possible DDM production channel: 
the pair production of heavy states which decay into four 
jets and a pair of DDM-ensemble constituents.
\label{fig:StrongProdDiagram}}
\end{figure}

In this section, we provide an example of a production process which
generically yields event counts of this order for any strongly-interacting 
parent particle with a TeV-scale mass: the pair-production of $\psi$,
as indicated in Fig.~\ref{fig:StrongProdDiagram}.  Indeed, any 
strongly-interacting exotic can be produced in this manner, via its 
interactions with the gluon field.  While the precise value 
of $N_e$ in any particular case depends on the chosen Lorentz and $SU(3)_c$ 
representations of $\psi$, 
on the value of $m_\psi$, and on the event-selection criteria imposed, we shall now 
demonstrate that event counts of the order assumed in Sect.~\ref{sec:significances} 
can easily be obtained with an integrated luminosity of 
roughly $\Lint \sim 30\mbox{~fb}^{-1}$ at the $\sqrt{s} = 14$~TeV LHC.

For any process via which $\psi$ is produced at colliders, the expected number 
of signal events $N_e$ surviving all cuts imposed on the data is given by
\begin{equation}
  N_e ~=~ N_\psi\sigma_{\psi}\Lint A\epsig \sum_n \mathrm{BR}(\psi\rightarrow jj\chi_n)~,
\end{equation}
where $\Lint$ is the integrated luminosity, $A$ is the detector acceptance,
$\epsig$ is the signal-event-selection efficiency, $\sigma_{\psi}$
is the cross-section for the production process in question, and $N_\psi$ is 
the multiplicity of $\psi$ particles produced in each event.  
For pair production, of course, $N_\psi = 2$.  Note that in 
assessing $\epsig$ for any particular production mechanism, we stipulate that
the corresponding event-selection efficiency for the SM background is such that
this background is negligible.  Moreover, in what follows, 
we shall assume for simplicity that the total branching fraction of $\psi$ to final 
states consisting of a pair of jets and one of the $\chi_n$ is effectively unity.
With these assumptions, the contribution to $N_e$ from pair production depends 
simply on the total pair-production cross-section $\sigma_{pp\rightarrow \psi\psi}$ 
for $\psi$, on the cuts imposed on the data (through $\epsig$), and on the inherent 
properties of the detector (through $A$). 

\begin{figure}[ht!]
\begin{center}
  \epsfxsize 3.0 truein \epsfbox {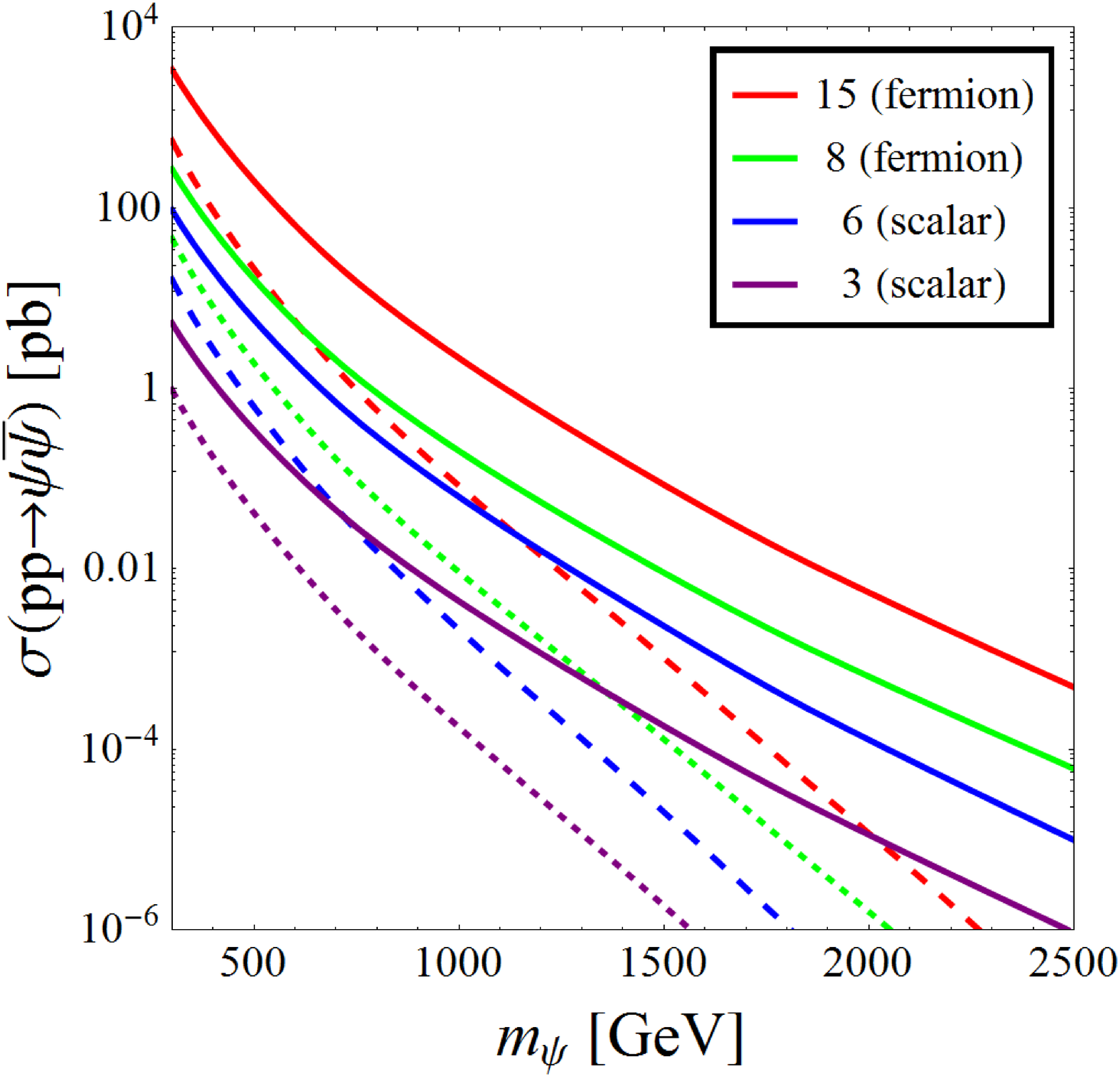}
  \epsfxsize 3.0 truein \epsfbox {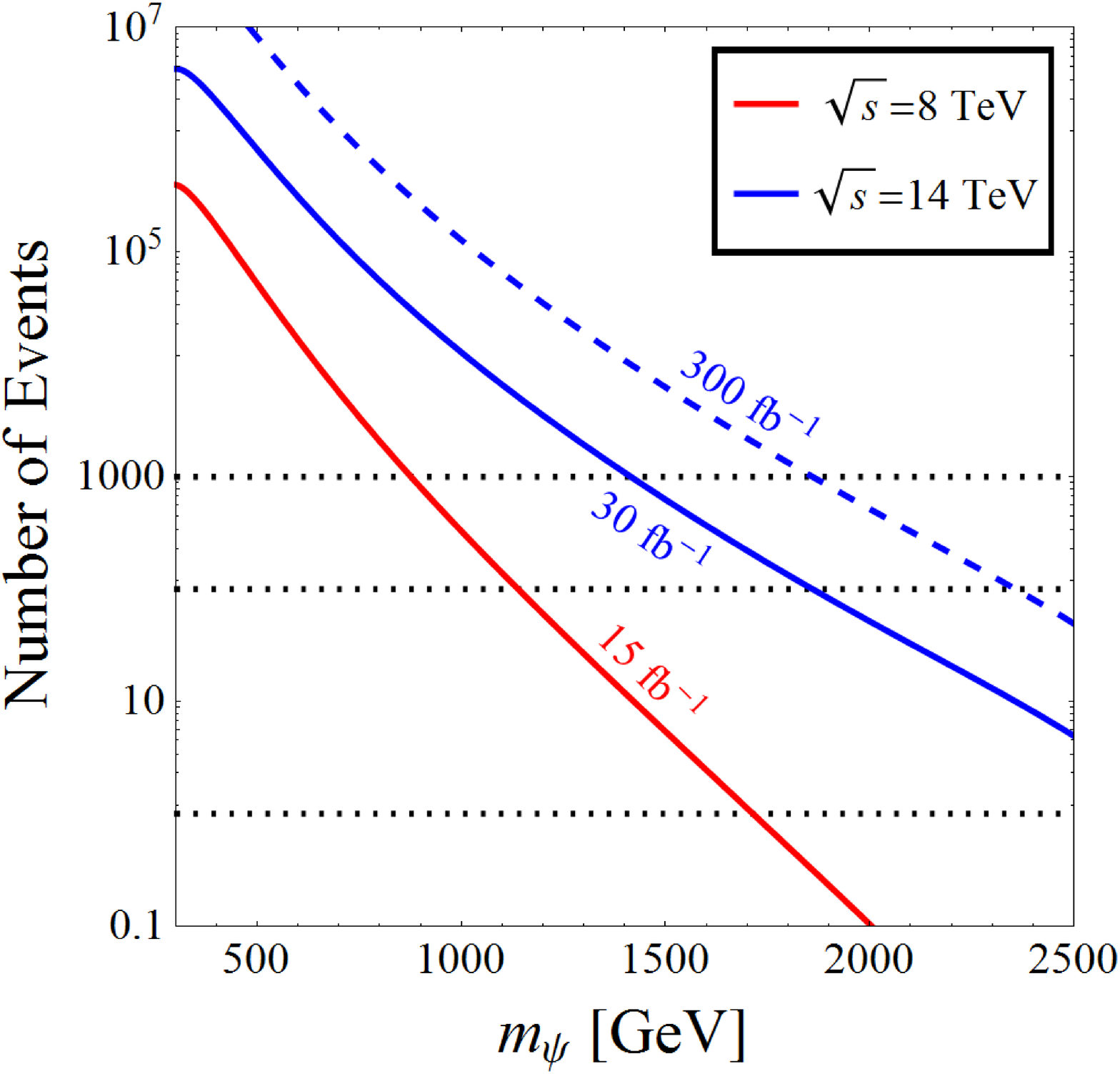} 
\end{center}
\caption{{\it Left panel}\/: The pair-production 
cross-sections $\sigma_{pp\rightarrow \psi\psi}$ obtained for a variety of 
relevant combinations of $SU(3)_c$ and Lorentz representations for a
strongly-interacting parent particle $\psi$.  The labels shown in the 
legend indicate the dimension of the $SU(3)_c$ representation for $\psi$ in 
each case, as well as whether $\psi$ is a scalar (as in the {\bf 3} and 
{\bf 6} cases shown), a Majorana fermion (as in the {\bf 8} case), or a 
Dirac fermion (as in the {\bf 15} case).  The dashed curves indicate 
the results for $\sqrt{s} = 8$~TeV, while the solid curves indicate the 
results for $\sqrt{s} = 14$~TeV.  
{\it Right panel}\/: the expected 
number of pair-production events $N_e$ (taking into account detector 
acceptance and signal efficiency) for a color-octet Majorana fermion 
$\psi$, plotted as a function of its mass $m_\psi$ for a variety of different 
integrated luminosities at the LHC.  The red curve corresponds to   
$\mathcal{L}_{\mathrm{int}}=15\mbox{~fb}^{-1}$ at $\sqrt{s} = 8$~TeV.  
The two blue curves correspond to 
$\mathcal{L}_{\mathrm{int}}=30\mbox{~fb}^{-1}$ (solid curve) and 
$\mathcal{L}_{\mathrm{int}}=300\mbox{~fb}^{-1}$ (dashed curve) 
at $\sqrt{s}=14$~TeV.  Dotted horizontal lines indicating $N_e = \{1,100,1000\}$
have also been included for convenience.  We thus see that significant values 
of $N_e$ can easily be achieved at the LHC through this production mechanism.
\label{fig:NEventsPlot}}
\end{figure}

We begin by evaluating $\sigma_{pp\rightarrow \psi\psi}$.
The parton-level pair-production cross-sections 
$\sigma_{\bar{q}q\rightarrow\psi\psi}$ and 
$\sigma_{gg\rightarrow\psi\psi}$ for an exotic scalar or fermion which 
transforms in an arbitrary representation of $SU(3)_c$ were calculated to leading 
order (LO) in Ref.~\cite{Taxation}.  Convolving these expressions with the appropriate 
parton-distribution functions (PDFs) in each case yields the LO hadronic 
cross-section $\sigma_{pp\rightarrow \psi\psi}$.  In this analysis, we use 
the CTEQ6L1~\cite{CTEQ6PDFs} PDF set, and impose a pseudorapidity cutoff $|\eta|\leq 3$ 
in computing the hadronic cross-sections in order to account for the finite 
pseudorapidity coverage of the LHC detectors.  The LO hadronic cross-sections
corresponding to several combinations of $SU(3)_c$ and Lorentz representations for 
$\psi$ for which decays of the form $\psi \rightarrow j j \chi_n$ can have sizeable 
branching fractions are shown in the left panel of Fig.~\ref{fig:NEventsPlot} as a 
function of $m_\psi$.  We see from this figure that in the case in which 
$\psi$ is a Majorana fermion transforming in the {\bf 8} representation
of $SU(3)_c$ or a Dirac fermion transforming in the {\bf 15} representation, 
a cross section in the range $\sigma_{pp\rightarrow \psi\psi}\gtrsim 100$~fb 
is attained for all values of $m_\psi \lesssim \{1100,1500\}$~GeV at the 
$\sqrt{s} = 14$~TeV LHC.  While the cross-sections are smaller for the 
cases in which $\psi$ is a scalar field transforming in the {\bf 3} or {\bf 6} 
representation of $SU(3)_c$, they are still be reasonably large for 
$m_\psi \lesssim 1000$~GeV.  

We now turn to address the values of $A$ and $\epsig$ associated with 
the pair-production of a strongly-interacting parent particle.
Since $\epsig$ depends sensitively on the details of the scenario in question and on
the particular set of event-selection criteria which provides an optimal 
discriminant between signal and background in that scenario, we focus on one 
representative example case: that in which $\psi$ is a color-octet Majorana 
fermion.  For this case, representative values for the product $A\epsig$ may 
be taken from results of searches for pair-produced gluinos by the 
ATLAS collaboration~\cite{ATLASJetsMET35pb,ATLASWebFigs} in the 
$\mathrm{jets}+\met$ channel 
in the regime in which all squarks are taken to be heavy. 
Specifically, we adopt the efficiencies obtained for a number $N_j \geq 3$ 
of high-$p_T$ jets and $m_{\mathrm{eff}} > 500$~GeV, where $m_{\mathrm{eff}}$ 
is the scalar sum of $\met$ and the magnitudes $p_{T_j}$ of the transverse
momenta of the three leading jets in the event, ranked by $p_T$.
In either case, $A\epsig$ lies within the range $0.3 - 0.6$ for 
$m_\psi \gtrsim 500$~GeV.  In addition, an estimate of the effects 
of next-to-leading-order (NLO) corrections to $\sigma_{pp\rightarrow\psi\psi}$ in this case can 
be obtained using the $K$-factor formalism, in which the LO cross-section is scaled 
by an overall multiplier.  The NLO $K$-factors for $\psi$ pair production used in 
this analysis were obtained using the Prospino package~\cite{Prospino}, and vary
from $K \approx 1.7 - 4.4$ for $300\lesssim m_\psi \lesssim 2500$~GeV at $\sqrt{s} = 8$~TeV
and from $K \approx 1.5 - 2.1$ at $\sqrt{s} = 14$~TeV for the same mass range. 

In the right panel of Fig.~\ref{fig:NEventsPlot}, we show the expected number 
$N_e$ of events at the ATLAS detector for the case in which $\psi$ 
is a color-octet Majorana fermion, plotted as a function of $m_\psi$ with 
$\Lint = 15\mbox{~fb}^{-1}$ for the current LHC run
at $\sqrt{s} = 8$~TeV (red), and with $\Lint = \{30,300\}\mbox{~fb}^{-1}$ for the 
projected upcoming run at $\sqrt{s} = 14$~TeV (blue).  
These results incorporate the effects of detector acceptance and signal 
efficiency, as discussed above, and include the relevant NLO $K$-factors.  
Dotted lines corresponding to $N_e = \{1,100,1000\}$ have also been included for 
reference.  Note that for this choice of representations, $m_\psi$ is constrained by
ATLAS gluino searches~\cite{ATLASJetsMET1fb} in the $\mathrm{jets}+\met$ channel 
in the context of supersymmetric models with heavy, decoupled squarks. 
Such searches likewise place a constraint $m_\psi \gtrsim 700$~GeV on the mass 
of any color-octet Majorana fermion which decays to a pair of jets and an 
invisible particle with a branching fraction near unity. 
      
The results displayed in the right panel of Fig.~\ref{fig:NEventsPlot} demonstrate
that the pair-production of strongly-interacting parent particles with TeV-scale 
masses is certainly capable of yielding $N_e \sim \mathcal{O}(500 - 1000)$ within the
first $30\mathrm{~fb}^{-1}$ of integrated luminosity at the $\sqrt{s} = 14$~TeV 
LHC --- an event count which was shown in Sect.~\ref{sec:significances} to be 
sufficient for conclusively
distinguishing DDM scenarios from more traditional dark-matter models on the basis 
of $\mij$ distributions.  Indeed, we see that $N_e \sim 1000$ can be expected for a 
color-octet fermion parent particle with $m_\psi \lesssim 1500$~GeV at such an 
integrated luminosity during the planned run at $\sqrt{s} = 14$~TeV.  
Furthermore, even with the integrated luminosity 
$\mathcal{L}_{\mathrm{int}}\approx 15\mathrm{fb}^{-1}$ anticipated by the conclusion 
of the current run at $\sqrt{s} = 8$~TeV, an event count of roughly 
$N_e \sim 1000$ is obtained for $m_\psi \lesssim 800$~GeV.  While acceptance 
and signal-efficiency factors will differ somewhat for parent particles with different 
spins and $SU(3)_c$ representations, the results shown in the left panel of
Fig.~\ref{fig:NEventsPlot} indicate that event counts of this order can
likewise be expected in a variety of other cases as well.  It should be noted,
however, that combinatorial ambiguities associated with the pairing of final-state jets 
frequently necessitate the imposition of additional cuts, and this can lead to a
reduction in signal efficiency.  In Sect.~\ref{sec:conclusions}, we shall discuss  
various methods through which such combinatorial ambiguities can be addressed. 


\section{Discussion and Conclusions\label{sec:conclusions}}


In this paper, we have examined the prospects for observing evidence of 
dynamical dark matter at colliders, and for distinguishing between 
DDM scenarios and more traditional dark-matter scenarios on the basis of 
collider data.  We have focused primarily on one promising technique: the 
identification of distinctive features imprinted on the kinematic   
distributions of SM fields produced along with the constituent fields of the 
DDM ensemble via the decays of other, heavier fields which happen to 
be present in a given DDM model.  We have shown that the distributions which arise in 
DDM models can differ substantially from the characteristic distributions 
which arise in traditional dark-matter models in which a single stable, 
neutral beyond-the-SM particle is responsible for essentially the entirety of $\OmegaDM$.  
To illustrate the efficacy of our approach, we have examined the prospects for 
observing a statistically significant deviation in the shape of the 
dijet invariant-mass distributions obtained for a particular representative 
DDM scenario relative to those obtained in traditional
dark-matter models.  We have demonstrated that throughout a 
substantial region of the parameter space of our scenario, a deviation at the 
$5\sigma$ level is obtained within the first $30\mbox{~fb}^{-1}$ of integrated
luminosity at the $\sqrt{s} = 14$~TeV LHC.  In particular, we have shown that 
regions of the parameter space of this scenario in which $\gamma > 0$ or in which
$\Delta m$ is small and $\delta < 1$ are particularly auspicious for 
differentiation.  These findings confirm that the LHC is capable of providing 
compelling evidence for dynamical dark matter within the next few years.           

We emphasize that the quantitative results we have obtained 
depend primarily on the total production rate for the parent particle and not on 
the particular manner in which this parent particle is produced.  Indeed, these results
apply directly to any situation in which the jets produced by the decay of each parent 
particle $\psi$ in the event can be unambiguously identified with that particular 
$\psi$.  In certain cases, such identification is indeed trivial.  For example, 
$\psi$ may be produced singly or have competing decays.  
Moreover, even in situations in which $\psi$ decays predominately via parton-level 
processes of the form $\psi\rightarrow \overline{q}q\chi_n$,
heavy-flavor tagging could assist in the identification of the correct pairings in 
events in which one parent particle decays to a final state involving 
light quarks and the other to a final state involving heavy 
quarks, \eg, $\overline{b}b$.
In many situations, however, it is not possible to identify the correct pairing
of final-state jets based on the properties of the jets alone, and 
incorrect jet pairings give rise to a substantial combinatorial background.
Nevertheless, a number of techniques exist for eliminating this 
combinatorial background.  These include the hemisphere
method~\cite{HemisphereMethod,FelixArvind}, 
the use of additional kinematic variables such as $m_{T_2}$~\cite{MT2}, 
and background-subtraction strategies~\cite{BEST}.  Indeed,    
it would be interesting to assess the relative efficacy of these techniques  
in helping to discriminate between DDM models and more traditional dark-matter 
scenarios in the presence of combinatorial ambiguities~\cite{DSRTfuture}.    

We note that we have chosen to focus on the case in which the SM
particles produced alongside the $\chi_n$ by $\psi$ decays are quarks or 
gluons solely because large event rates can generically be expected for 
strongly-interacting $\psi$.  However, the techniques that we have developed here
can also be applied to situations in which $\psi$ decays to final states 
involving different combinations of SM particles as well.    
For example, the prospects for differentiating between
DDM scenarios and traditional dark-matter models obtained for 
decay topologies such as $\psi \rightarrow \ell^+\ell^-\chi_n$, where 
$\ell^\pm$ denotes a charged lepton, will be identical to those obtained 
for $\psi \rightarrow j j\chi_n$, given the same total number of events $N_e$.  

We also note that the techniques outlined here for differentiating between DDM 
ensembles and more traditional dark-matter candidates also have a range of 
applicability that extends beyond the DDM framework.  
Indeed, not every particle which manifests itself as $\met$ at colliders 
necessarily contributes significantly (or at all) to $\OmegaDM$, and $\met$
signatures of this sort can arise in a number of theories involving multiple neutral 
particles which are stable on collider time scales but not necessarily stable on time 
scales approaching the age of the universe.  As discussed in Sect.~\ref{sec:production}, 
examples of such theories can be found among supersymmetric models with highly 
compressed spectra~\cite{CompressedSUSY}, scenarios involving universal extra 
dimensions~\cite{Antoniadis,DDGLargeED,UED}, and a wide variety of additional 
extensions of the SM.

Finally, we note that in this paper we have focused on the case in which 
$\psi$ decays to a final state consisting of SM states and one or more of the 
$\chi_n$ directly, via an effective multi-body interaction vertex.  
Alternatively, one could also consider the situation
in which $\psi$ decays to such a final state primarily via decay 
chains involving on-shell intermediate states.  Indeed, the kinematic distributions
associated with the final-state SM fields in this latter case differ significantly 
from those obtained in the former.  For example, in traditional dark-matter models, 
the $\mij$ distributions associated with $\psi\rightarrow j j\chi$ processes
which proceed via cascade decays involving an on-shell intermediary take a 
characteristic triangular shape.  It would be interesting to examine the 
corresponding $\mij$ distributions which arise for DDM ensembles, and whether 
such dark-matter candidates can likewise conclusively be distinguished from 
traditional dark-matter candidates on the basis of these 
distributions~\cite{DSTfuture}.
        

\begin{acknowledgments}


We would like to thank Z.~Chacko, J.~Rutherfoord, S.~Vahsen, and E.~Varnes for discussions. 
KRD and SS are supported in part by the U.S. Department of Energy under Grant 
DE-FG02-04ER-41298, and KRD is additionally supported in part by the National 
Science Foundation through its employee IR/D program.
BT is supported in part by DOE grant DE-FG02-04ER-41291.  The opinions and 
conclusions expressed here are those of the authors, and do not represent either 
the Department of Energy or the National Science Foundation.

\end{acknowledgments}

\bigskip


\end{document}